\pdfoutput=1
\documentclass[11pt]{article}
\usepackage{jheppub}

\usepackage{amsmath}
\usepackage{amssymb}
\usepackage{graphicx,color}
\usepackage{amsthm}
\usepackage{subfigure}

\usepackage{longtable}

\usepackage{ifpdf}

\usepackage[boxsize=2em,aligntableaux=center]{ytableau}

\usepackage{multirow}

\allowdisplaybreaks

\newcommand{\be}{\begin{equation}}
\newcommand{\ee}{\end{equation}}
\newcommand{\ba}{\begin{eqnarray}}
\newcommand{\ea}{\end{eqnarray}}
\newcommand{\nn}{\nonumber}
\newcommand{\lp}{\left(}
\newcommand{\rp}{\right)}

\newcommand{\w}{\wedge}

\newcommand{\N}{\mathcal{N}}
\newcommand{\bF}{\mathbb{F}}
\newcommand{\bZ}{\mathbb{Z}}
\newcommand{\cZ}{\mathcal{Z}}

\notoc

\title{\centering Comments on M$_{24}$ representations and $CY_3$ geometries}

\author[]{Natalie M. Paquette,}
\author[]{Timm Wrase}

\affiliation[]{Stanford Institute for Theoretical Physics\\ Stanford University, Stanford, CA 94305, USA}

\emailAdd{npaquett@stanford.edu}
\emailAdd{timm.wrase@stanford.edu}

\abstract{We show using string dualities that Mathieu moonshine controls Gromov-Witten invariants \emph{and} periods of the holomorphic 3-form $\Omega$ for certain $CY_3$ manifolds. We also discuss how the period vectors appear in flux compactifications on these $CY_3$ manifolds and work out the connection between the sporadic group M$_{24}$ and the Yukawa couplings in four dimensional theories that arise from heterotic string theory compactifications on these $CY_3$ manifolds.
}

\begin{document}

\makeatletter
\let\old@fpheader\@fpheader
\renewcommand{\@fpheader}{\old@fpheader\hfill
SU/ITP-14/19}
\makeatother

\maketitle

\newpage

\section{Introduction}
In 2010 Eguchi, Ooguri and Tachikawa \cite{EOT} showed that the elliptic genus of the $K3$ manifold can be expanded in such a way that the first few expansion coefficients are sums of dimensions of irreducible representations of the largest Mathieu group M$_{24}$. This connection between the elliptic genus of $K3$ and M$_{24}$ was checked and confirmed in \cite{Miranda, Gaberdiel:2010ch, Gaberdiel:2010ca, Eguchi:2010fg}.\footnote{For very interesting generalizations of this moonshine see \cite{Cheng:2012tq, Cheng:2013wca, Cheng:2014zpa} and \cite{Persson:2013xpa}.} In 2012, Gannon proved \cite{Gannonproof} that all the expansion coefficients appearing in the elliptic genus are sums of irreducible representations of M$_{24}$. Despite all this work, there are still many interesting questions related to this `Mathieu moonshine' that have not yet been answered. For example, no $\N=(4,4)$ non-linear sigma model with $K3$ target has M$_{24}$ as its symmetry group \cite{Gaberdiel}. So why does the elliptic genus of $K3$ exhibit this connection to M$_{24}$? One possible explanation, currently pursued in, for example, \cite{Taormina:2011rr, Taormina:2013jza}, is that the symmetries of different points in $K3$ moduli space combine to give M$_{24}$. An alternative idea is that models which preserve only $\N=(0,4)$ worldsheet supersymmetry and that are connected to $\N=(4,4)$ non-linear sigma model with $K3$ target, have as their symmetry group the full M$_{24}$ group \cite{Cheng:2013kpa, Harrison:2013bya}.

Since Mathieu moonshine involves the $K3$ manifold that has played a major role in compactifications of superstring theories and in string dualities, it is very interesting for string theorists. We are currently in the process of understanding the implications of this moonshine phenomenon for superstring compactifications and have already obtained a variety of new insights: For example, it was shown in \cite{Hohenegger:2011us} that certain one-loop amplitudes in compactifications of type II string theory on $K3 \times T^2$ are related to the elliptic genus of $K3$ and therefore to Mathieu moonshine. In \cite{Harvey:2013mda}, the authors found that certain BPS states in type II string theory compactified on $S^1 \times K3$ are related to a particular mock modular form that is closely related to the elliptic genus of $K3$. Compactifying the heterotic string theory on $K3 \times T^2$, the authors of \cite{Cheng:2013kpa} showed that the sums of irreducible representations of M$_{24}$ that appear in Mathieu moonshine also appear (albeit in a less direct manner) in the prepotential of the resulting four dimensional $\N=2$ theories. \footnote{In the case of the standard embedding, where there exists a $(4, 4)$ locus in the $(0, 4)$ moduli space, it is perhaps reasonable to decompose the prepotential into $\mathcal{N}=4$ characters to observe the appearance of M$_{24}$ representations. This would correspond on the type IIA side to a compactification on the threefold with base $\mathbb{F}_{12}$. However, it is unclear why the $\mathcal{N}=4$ characters, rather than e.g. Virasoro characters augmented by a $U(1)$ current algebra, continue to work for other embeddings. It would be interesting to understand this point better; for now, we can simply say that the $\mathcal{N}=4$ decompositions, perhaps miraculously, work.} To support the conjecture that Mathieu moonshine plays a role in these $\mathcal{N}=2$ compactifications, a variety of twined elliptic genera (i.e. the analogue of the McKay-Thompson series for the Monster) were calculated in \cite{Harrison:2013bya}, in which the authors twined by explicit symmetries of heterotic GLSMs with $K3$ target, for various instanton embeddings. For some of these symmetries, the twined elliptic genera reproduced the graded traces predicted by Mathieu moonshine. These heterotic theories are dual to type IIA compactifications on $CY_3$ manifolds $X_n$ that are elliptic fibrations over the Hirzebruch surfaces $\mathbb{F}_n$ for $n=0,1,\ldots,12$. In these dual type IIA theories the prepotential receives instanton corrections and those are by duality related to the Mathieu group M$_{24}$ \cite{Cheng:2013kpa}. More specifically, the instanton corrections are determined by the Gromov-Witten invariants of the $CY_3$ manifolds $X_n$ and these are connected to Mathieu moonshine. This extends the usual connection between number theory and representation theory that is heralded by the appearance of moonshine to also include (algebraic) geometry. Furthermore, the corrections to the prepotential determine the gauge couplings in the four-dimensional $\N=2$ spacetime theories. Hence, the 1-loop corrections to the gauge couplings are implicated in Mathieu moonshine. Such a connection appears more generally in heterotic string theory compactifications. It was shown in \cite{Wrase:2014fja} that for almost all four-dimensional $\N=1$ theories that arise from heterotic orbifold compactifications, the gauge kinetic functions (and therefore the gauge couplings) receive a universal one-loop correction that is connected to the Mathieu group M$_{24}$.

We see that Mathieu moonshine has already lead to a variety of intriguing new insights for several different compactifications of superstring theories. In this paper we add to this list by applying mirror symmetry to the above type IIA compactifications on the $CY_3$ manifolds $X_n$ that are elliptic fibrations over the Hirzebruch surfaces $\mathbb{F}_n$. Mirror symmetry relates the Gromov-Witten invariants of $X_n$ to the periods of the holomorphic 3-form $\Omega$ of the mirror $Y_n$. We explicitly work out the connection between these periods and representations of M$_{24}$ for $Y_n$ with $n=0,1,2$, though the results will generalize to all $n$ in an obvious way. Having implicated the holomorphic 3-forms of the $Y_n$ in Mathieu moonshine, we note that for $n=2,4,6,12$, the $X_n$ are given as hypersurfaces in the weighted projective space $\mathbb{WP}_{1,1,n,2n+4,3n+6}$ and the mirror manifolds $Y_n$ can be obtained from a Greene-Plesser type construction \cite{greene1990duality}. This means that one expects that a subspace of the complex structure moduli space of these particular $X_n$ is the same as the complex structure moduli space of the $Y_n$ (and likewise for the quantum K\"ahler moduli space). For at least $n=2,4,6,12$ there would then be a connection between M$_{24}$ and the K\"ahler as well as the complex structure sector of the $X_n$ and $Y_n$. Having established such a link, we then proceed and discuss two implications for physically interesting theories. First we study flux compactifications on $X_n$ and $Y_n$ and show how M$_{24}$ representations appear in the Gukov-Vafa-Witten superpotential. Then we discuss compactification of the heterotic $E_8 \times E_8$ string theory on $X_n$ and $Y_n$ and find that the Yukawa couplings and therefore the masses of the particles in the resulting four-dimensional $\N=1$ theories are implicated in Mathieu moonshine as well.

The outline of the paper is as follows: In section \ref{sec:Moonshine}, we review Mathieu moonshine and show how through string dualities it controls Gromov-Witten invariants \emph{or} periods of the holomorphic 3-form $\Omega$ for certain $CY_3$ manifolds. Then we argue in section \ref{sec:M24both} that at least for some $CY_3$ manifolds the complex structure \emph{and} K\"ahler moduli space is implicated in Mathieu moonshine. Next we study flux compactifications on these manifolds in section \ref{sec:fluxcomp} and explicitly show how M$_{24}$ representations appear in the superpotential. In section \ref{sec:Yukawa} we show for certain compactifications of the heterotic string theory, how the Yukawa couplings of the 4d $\N=1$ theories are related to M$_{24}$. We summarize our findings and point out interesting future directions in section \ref{sec:conclusion}. Appendix \ref{app:mirrorsym} provides a concise introduction to mirror symmetry and appendix \ref{app:data} lists topological data for three $CY_3$ manifolds that are of particular interest to us.

\section{Mathieu Moonshine and the holomorphic 3-form $\Omega$}\label{sec:Moonshine}
In this section we first review Mathieu moonshine that was discovered in \cite{EOT}. There the authors expand the elliptic genus of the $K3$ manifold and find that the expansion coefficients are sums of dimensions of irreducible representations of the largest Mathieu group M$_{24}$. Then we use the duality between heterotic string theory compactifications on $K3 \times T^2$ and type IIA compactifications on $CY_3$ manifolds $X_n$ that are elliptic fibrations over $\mathbb{F}_n$ to discuss (following \cite{Cheng:2013kpa}) how Mathieu moonshine is connected to the Gromov-Witten invariants of the $X_n$. Using mirror symmetry we finally connect Mathieu moonshine to the holomorphic 3-form $\Omega$ of $Y_n$, that are the mirror $CY_3$ manifolds of the $X_n$. We then argue using the Greene-Plesser construction of mirror pairs that at least some of $X_n$ and $Y_n$ exhibit a connection between M$_{24}$ and \emph{both} their Gromov-Witten invariants and their holomorphic 3-form $\Omega$.

\subsection{Mathieu moonshine}
The elliptic genus is defined as the following trace over the RR sector of an $\N=(2,2)$ superconformal field theory
\be
\cZ_{elliptic}(q,y) = \text{Tr}_{RR} \lp (-1)^{F_L+F_R} q^{L_0-\frac{c}{24}} y^{J_0} \bar{q}^{\bar{L}_0-\frac{\bar{c}}{24}} \rp\,.
\ee
Here $F_{L/R}$ denotes the left/right moving fermion number and $y$ is a chemical potential for the left-moving $U(1)$ charge measured by $J_0$. Since only the right-moving Witten index $(-1)^{F_R} \bar{q}^{\bar{L}_0-\frac{\bar{c}}{24}}$ appears in $\cZ_{elliptic}$, it does not depend on $\bar{q}$. For the particular case of $K3$, the elliptic genus was calculated in 1989 in \cite{Eguchi:1988vra}. It wasn't until 2010, however, that Eguchi, Ooguri and Tachikawa \cite{EOT} noticed that the coefficients appearing in the $K3$ elliptic genus expanded in terms of $\N=4$ characters are related to the dimensions of irreducible representations of M$_{24}$. In particular, if we define the $\N=4$ superconformal characters \cite{Eguchi:1988vra}  (please see appendix A of \cite{Cheng:2013kpa} for our conventions for the Jacobi $\theta$-functions)
\ba
\text{ch}_{h=\frac14,l=0}(q,y) &=& -\frac{i y^{\frac12} \theta_1(q,y)}{\eta(q)^3} \sum_{n=-\infty}^{\infty} \frac{(-1)^n q^{\frac12 n(n+1)} y^n}{1-y \ q^n}\,,\\ \text{ch}_{h=n+\frac14,l=\frac12}(q,y) &=& q^{n-\frac18}\frac{\theta_1(q,y)^2}{\eta(q)^3}\,,
\ea
then one finds the following expansion \cite{EOT}
\ba
\cZ_{elliptic}^{K3}(q,y) &=& 8 \left[ \lp \frac{\theta_2(q,y)}{\theta_2(q,1)}\rp^2 + \lp \frac{\theta_3(q,y)}{\theta_3(q,1)}\rp^2 + \lp \frac{\theta_4(q,y)}{\theta_4(q,1)}\rp^2 \right]\\
&=& 24 \ \text{ch}_{h=\frac14,l=0}(q,y) + \sum_{n=0}^\infty A_n \text{ch}_{h=n+\frac14,l=\frac12}(q,y)\,.
\ea
The 24 = 23+1 as well as the first few $A_n$ where identified in \cite{EOT} as sums of irreducible representations of M$_{24}$
\ba\label{eq:As}
A_0 &=& -2=-1-1\,,\nn\\
A_1 &=& 90 = 45 + \overline{45}\,,\nn\\
A_2 &=& 462 = 231 + \overline{231}\,,\\
A_3 &=& 1540 = 770 + \overline{770}\,,\nn\\
A_4 &=& 4554 = 2277+2277\,,\nn\\
&\ldots&
\ea
It was proven in \cite{Gannonproof} that all the $A_n$ for $n\geq1$ are sums of irreducible representations of M$_{24}$ with only positive coefficients.

\subsection{M$_{24}$ in Type II $\mathcal{N}=2$ theories}

This connection between the elliptic genus of $K3$ and the Mathieu group M$_{24}$ is still not understood and one might hope that studying the appearance of this Mathieu moonshine in different string theory settings might help understand it better. In addition, this might lead to new insights in otherwise well-understood string compactifications and connections between physical observables and the Mathieu group M$_{24}$ in certain toy models. Of particular interest to us is \cite{Cheng:2013kpa}, where it was shown that the elliptic genus of $K3$ appears in compactifications of the heterotic string theory and that, by duality, the Gromov-Witten invariants of certain $CY_3$ manifolds are related to the Mathieu group M$_{24}$. After quickly reviewing these results we will extend them and show explicitly how the holomorphic three form of certain $CY_3$ manifolds is related to M$_{24}$.

The heterotic $E_8 \times E_8$ string theory compactified on $K3 \times T^2$ leads to a four dimensional spacetime theory with $\N=2$ supersymmetry (see for example \cite{Lusttwo} for a nice review of basic facts about these theories). In order to satisfy the Bianchi identity for the $H_3$ field one has to turn on a non-trivial gauge bundle inside one or both of the $E_8$ gauge groups. In particular, in the absence of NS5-branes, we have to embed a total of 24 instantons into the two $E_8$'s which leads to 13 different cases due to the symmetry that exchanges the two $E_8$'s. We embed $(12-n,12+n)$ instantons in $E_8 \times E_8$ and take w.l.o.g. $n=0,1,\ldots,12$. These 13 cases are perturbatively inequivalent, however, each case can be further subdivided based on the particular subgroup $G \times G' \subset E_8 \times E_8$ in which one turns on the instantons.

For $n=0,1,2$ the instantons generically break the $E_8 \times E_8$ gauge symmetry and there are only three vector multiplets whose scalar components we denote by $S, T$ and $U$. $S$ is the axio-dilaton, while $T$ and $U$ control the size and complex structure of the two torus $T^2$. For $n>2$ there are additional Wilson line moduli $V^i$. As was shown in \cite{Cheng:2013kpa}, after setting the Wilson line moduli to zero $V^i=0$, the prepotential for the thirteen four dimensional $\N=2$ spacetime theories is always the same and is directly related to the elliptic genus of $K3$ and therefore to M$_{24}$.\footnote{If one embeds all instantons in one $E_8$ and allows for non-zero Wilson lines for the other $E_8$, then there is still a direct connection between the prepotential and M$_{24}$ \cite{Cheng:2013kpa}.} In particular (up to a quadratic polynomial in $S$, $T$ and $U$) it is given by
\be\label{eq:prepot}
F = STU + \frac{1}{3}U^3 +\frac{1}{(2\pi i)^3} c(0) \zeta(3) -\frac{2}{(2\pi i)^3} \sum_{\substack{k>0, l \in \mathbb{Z} \\ k=0,l>0}} c(kl) Li_3\lp q_T^k q_U^l \rp + \mathcal{O}(e^{2\pi i S})\,,
\ee
where $\zeta(3)\approx 1.2$ is the Riemann zeta function, $q_U = e^{2 \pi i U}$, $q_T = e^{2 \pi i T}$, the polylogarithm is defined as $Li_p = \sum_{n=1}^\infty \frac{x^n}{n^p}$ and the coefficients $c(m)$ are obtained from the expansion
\be\label{eq:cs}
 \frac{E_4(q) E_6(q)}{\eta(q)^{24}} = \sum_{m \geq -1} c(m) q^m = \frac{1}{q} - 240 - 141444 q -\ldots\,, \quad \text{and} \quad c(m)=0\quad \forall \ m <-1,
\ee
where $E_i(q)$ are the Eisenstein series (see appendix A in \cite{Cheng:2013kpa} for a definition).

From the explicit derivation of the prepotential one finds that $E_6(q)$ and therefore the $c(m)$ in the prepotential \eqref{eq:prepot} are related to the elliptic genus of $K3$. Explicitly one has
\ba
-\frac{4 E_6(q)}{\eta(q)^{12}} &=& \lp \frac{\theta_2(q)}{\eta(q)} \rp^6  \ \cZ_{elliptic}^{K3}(q,-1) + \lp \frac{\theta_3(q)}{\eta(q)} \rp^6  \ q^{\frac14} \ \cZ_{elliptic}^{K3}(q,-q^{\frac12}) \nn\\
&& - \lp \frac{\theta_4(q)}{\eta(q)} \rp^6  \ q^{\frac14} \ \cZ_{elliptic}^{K3}(q,q^{\frac12})\nn\\
&=& 24 g_{h=\frac14,l=0}(q) + g_{h=\frac14,l=\frac12}(q) \sum_{n=0}^\infty A_n q^n\,,
\ea
where the 24=23+1 and the $A_n$'s can be decomposed into irreps of M$_{24}$ as in \eqref{eq:As} and we defined
\ba
g_{h=\frac14,l}(q) &=& \lp \frac{\theta_2(q)}{\eta(q)} \rp^6 \text{ch}_{h=\frac14,l}(q,-1) + \lp \frac{\theta_3(q)}{\eta(q)} \rp^6  \ q^{\frac14} \ \text{ch}_{h=\frac14,l}(q,-q^{\frac12}) \nn\\
&& - \lp \frac{\theta_4(q)}{\eta(q)} \rp^6  \ q^{\frac14} \ \text{ch}_{h=\frac14,l}(q,q^{\frac12})\,.
\ea
Having established this connection between the Mathieu group M$_{24}$ and the $\N=2$ prepotential in the spacetime theory, the authors of \cite{Cheng:2013kpa} used the fact that these compactifications of the heterotic $E_8 \times E_8$ string theory are dual to compactifications of type IIA on $CY_3$ manifolds $X_n$ that are elliptic fibrations over the Hirzebruch surfaces $\mathbb{F}_n$, where again $n=0,1,\ldots 12$.\footnote{We group together all the $CY_3$ manifolds that are dual to heterotic constructions with the same instanton numbers and collectively call them $X_n$. All manifolds for a given $n$ are related by geometric transitions that correspond to (un-)higgsing the gauge group on the dual heterotic side.} In the dual type IIA compactification the infinite sum in the prepotential \eqref{eq:prepot} arises from instanton corrections and the $c(m)$ are related to the Gromov-Witten invariants of the $CY_3$ manifolds $X_n$. The prepotential on the type IIA side was recently calculated in \cite{Klemm, Scheidegger} for $X_0, X_1$ and $X_2$ and it matches the heterotic result \eqref{eq:prepot} to leading order in $q_T$ \cite{Cheng:2013kpa}. Thus, there is a connection between Gromov-Witten invariants of certain $CY_3$ manifolds and the sporadic group M$_{24}$.

We now review that by mirror symmetry this implies that for certain $CY_3$ manifolds the holomorphic 3-form $\Omega$ is likewise connected to the Mathieu group M$_{24}$. Mirror symmetry, as we review in appendix \ref{app:mirrorsym}, is a duality between compactifications of type IIA string theory on a Calabi-Yau manifold $X_n$ and type IIB string theory on the mirror Calabi-Yau manifold $Y_n$. The moduli space of four dimensional $\N=2$ theories (locally) factorizes into a hypermultiplet part and a vector multiplet part. In our particular compactifications of the heterotic and type IIA string theories the vector multiplets are connected to M$_{24}$. In compactifications of type IIA string theory the vector multiplets arise from the K\"ahler moduli sector, while for the dual type IIB string theory compactifications the vector multiplets arise from the complex structure sector. So we expect that the mirror $CY_3$ manifolds $Y_n$ have a complex structure moduli space that is related to M$_{24}$.

In particular, as discussed in appendix \ref{app:flatcoord}, we can integrate the holomorphic three form $\Omega$ of the $CY_3$ manifolds $Y_n$ over a canonical homology basis such that
\be
z^i = \int_{A^i} \Omega\,, \qquad \mathcal{F}_i(z) = \int_{B_i} \Omega\,.
\ee
In the basis of forms dual to $\left \lbrace A^i, B_i \right \rbrace$, it is often convenient to expand the 3-form as $\Omega= z^i \alpha_i - \mathcal{F}_i(z) \beta^i$. As we will explain, the $\mathcal{F}_i$ exhibit interesting dependence on M$_{24}$ via their dependence on the holomorphic prepotential $\mathcal{F}$: $\mathcal{F}_i = \partial_{z^i} \mathcal{F}$.

The prepotential that controls the vector multiplet moduli space for type IIB compactifications on $Y_n$ is given by $\mathcal{F} = \frac12 z^i \mathcal{F}_i(z)$, which is a function of the projective coordinates $z^i$.   The periods themselves are solutions of the Picard-Fuchs equations, which can be determined with the classical intersection numbers of the mirror, $X_n$, as input. The mirror map can also be inferred from the solutions to the Picard-Fuchs equations in an expansion around $z^i = 0$. This large complex structure point is mirror to the large radius point of $X_n$, so applying the mirror map in an expansion around this point enables us to read off the Gromov-Witten invariants of $X_n$. Therefore, the period vector of $Y_n$ is controlled entirely by some classical topological numbers plus the Gromov-Witten invariants of its mirror $X_n$.

The Gromov-Witten invariants come from the worldsheet instanton corrections to the K\"ahler moduli space of $X_n$, which must be small for our perturbative expansion to be valid. It is important to remember, though, that the complex structure moduli space of $Y_n$ is classically exact and its periods are expressible in a simple closed form in the $z^i$ coordinates.

The period vector can be expressed in terms of the prepotential as (see Appendix \ref{app:mirrorsym} for details):
\be
\Pi =
\begin{pmatrix}
1 \\
t^i \\
\frac{\partial}{\partial t^i}\frac{\mathcal{F}}{(z^0)^2} \\
2 \frac{ \mathcal{F}}{(z^0)^2} - t^i \frac{\partial}{\partial t^i}\frac{ \mathcal{F}}{(z^0)^2}
\end{pmatrix}\,,
\ee
where $t^i=z^i/z^0$ are the three moduli dual to $S, T, U$ on the heterotic side. Since the Hirzebruch surfaces may be viewed as certain $\mathbb{P}^1$ fibrations over $\mathbb{P}^1$, the $t^i$ measure the volumes of the elliptic fiber and the two $\mathbb{P}^1$s: $t^i = \int_{C_i} (B + i J)$, where $B$ is the NS-NS field and $J$ is the K\"ahler form.\footnote{In the context of the type II string, we may view our compactification manifold  as being either an elliptic fibration over $\mathbb{F}_n$ or a $K3$ fibration over $\mathbb{P}^1$. The elliptic fibration over $\mathbb{F}_2$, which we will study extensively in the next section, is a hypersurface $X_{24}(1, 1, 2, 8, 12)$ in a weighted projective space. The $K3$ fiber of the latter point of view is a  hypersurface in $X_{12}(1, 1, 4, 6)$.}
Finally, we can write the period vector even more explicitly by plugging in $F:= \mathcal{F}/(z^0)^2$
\be
F= \left(\frac{\kappa^{0}_{ijk}}{6} t^i t^j t^k + \frac12 a_{ij} t^i t^j + b^i t^i + \frac{\chi(X_n) \zeta(3)}{2 (2\pi i)^3} + \frac{1}{(2 \pi i)^3} \sum_{(n_i)} N_{(n_i)} Li_{3}(q_i^{(n_i)}) \right)\,.
\ee
The $\kappa^{0}_{ijk}$ are the classical triple intersection numbers of $X_n$. $a_{ij}$ and $b_i$ are also classical topological numbers which we define in Appendix \ref{app:mirrorsym}. We list their numerical values for $X_0, X_1$ and $X_2$ in Appendix \ref{app:data}. The $N_{(n_i)}$ are the Gromov-Witten invariants of $X_n$, of which at least a subset is governed by Mathieu moonshine, as we will delineate shortly.

With the aforementioned substitution, the period vector becomes:
\be
\Pi =
\begin{pmatrix}\label{eq:pervector}
1 \\
t^i \\
\frac{\kappa^{0}_{ijk}}{2} t^j t^k + a_{ij}t^j + b^i + \partial_{t^i}\left(F_{inst} \right) \\
-\frac{\kappa^{0}_{ijk}}{6} t^i t^j t^k + b_i t^i + c + 2 F_{inst} - t^i \partial_{t^i} F_{inst}
\end{pmatrix}\,,
\ee
where we have defined $F_{inst} = \frac{1}{(2 \pi i)^3} \sum_{(n_i)} N_{(n_i)} Li_{3}(q_i^{(n_i)})$.
In practice, it is easiest to compute the prepotential (in the $t^i$ coordinates, expanded around the large complex structure/large radius point) and the Gromov-Witten invariants directly from \eqref{eq:pervector} or by computing a triple integral of $\kappa_{ijk}[X_n]$ (see Appendix \ref{app:mirrorsym}) with the classical topological numbers as input. This is what we have done; we record the $\kappa_{ijk}[X_n] = \bar{\kappa}_{ijk}[Y_n]$ for $n=0,1,2$ to fifth order in the $q_i = e^{2 \pi i t^i}$ in Appendix \ref{app:data}.

Finally, we wish to verify that our mirror symmetry computations exhibit the moonshine that we expect from the heterotic/IIA duality described earlier. After computing the prepotential, we finally have all the necessary information in hand. First, we note that the duality is good on the heterotic side when the string coupling is small. This corresponds to $S \rightarrow 0$, which for us means ``ignoring" instanton contributions from what in the notation of Appendix \ref{app:data} we call $q_2$ in the elliptic fibration over $\mathbb{F}_0$ and $q_3$ for $\mathbb{F}_{1, 2}$. We simply use the usual type IIA/heterotic dictionary \cite{Lusttwo} and match
\be\label{eq:matching}
-2 c_{STU}(k l ) =N_{\text{II}}(k+l,0, k)[X_{0}]= N_{\text{II}}(k+l, k, 0)[X_{1,2}]\,,
\ee
which are the coefficients of the $F_{inst}$ on each side of the duality.\footnote{The notation $N_{\text{II}}(k + l, k, 0)$ indicates that we are looking at terms in the instanton expansion of order $Li_3(q_1^{k+ l} q_2^k q_3^0)$.}

In \cite{Cheng:2013kpa} \eqref{eq:matching} was explicitly checked for $k=1$. We have calculated the $N_{\text{II}}$, now allowing both $k$ and $l$ to vary, to $20^{th}$ order for each threefold and recovered the coefficients of $-2 E_4(q) E_6(q)/\eta(q)^{24}$, which exhibit M$_{24}$ moonshine, as expected from \eqref{eq:matching}. This constitutes a new numerical check of the duality at higher instanton number in the K3 fibers. We see explicitly that the connection to the M$_{24}$ persists when both the K3's elliptic fiber and $\mathbb{P}^1$ base are ``counted" multiple times.

The presence of $-2 E_4(q) E_6(q)/\eta(q)^{24}$ in the STU model and its corresponding influence on the IIA side have been known for a long time. The first mirror symmetry computations of this type were done in \cite{hosono1995mirror}, where the first few such Gromov-Witten invariants for $X_2$ were computed. As we see from our computations, these coefficients are also visible in the other $X_n$, indicating that the new connection to M$_{24}$ is indeed independent of the instanton embedding on the heterotic side. We emphasize that the $S \rightarrow 0$ limit corresponds to a large base $\mathbb{P}^1$ on the IIA side, so the Gromov-Witten invariants relevant for moonshine come from worldsheet instantons mapping into the $K3$ fiber. This seemingly different connection between $K3$ and M$_{24}$ certainly deserves further study and we point out in the conclusion that it could potentially extend to other $K3$ fibered $CY_3$ manifolds.

Having established the relationship between the sporadic group M$_{24}$ and the Gromov-Witten invariants of the $CY_3$ manifolds $X_n$, as well as the holomorphic 3-form $\Omega$ of the mirror manifolds $Y_n$, we show in the next section that for (at least some of) the $X_n$ part of the complex structure moduli space is also linked to M$_{24}$, and likewise for part of the K\"ahler moduli space of (at least some of) the $Y_n$.  We also discuss which physical implications can be derived from such a connection. Here we mostly focus on the holomorphic 3-form $\Omega$ of the $Y_n$ (and some of the $X_n$) and show in the section \ref{sec:fluxcomp} that its relation to M$_{24}$ leads to the appearance of dimension of M$_{24}$ in the Gukov-Vafa-Witten \cite{Gukov:1999ya} flux superpotential. In section \ref{sec:Yukawa}, we show that for compactifications of the heterotic string theory on the $X_n$ or $Y_n$  the Yukawa couplings of the four dimensional $\N=1$ theories are related to M$_{24}$.

\section{Connecting both complex structure \emph{and} K\"ahler moduli spaces to M$_{24}$}\label{sec:M24both}
For $n=2,4,6,8,12$ we can write the $X_n$ as hypersurfaces in the weighted projective space $\mathbb{WP}_{1,1,n,2n+4,3n+6}$. For at least $n=2,4,6,12$ the mirror manifolds can be obtained from a Greene-Plesser construction, because the sum of the weights is divisible by each weight (see \cite{greene1990duality}). This means that we can quotient the space $X_n$ by the maximal group of scaling symmetries to get a singular limit of its mirror, the $Y_n$ manifold.

For example, for $X_2$, the elliptic fibration over $\mathbb{F}_2$, we have the Hodge numbers $h^{1,1}=3$ and $h^{2,1}=243$, where the three K\"ahler moduli correspond to the three $STU$ moduli of the previous section. If we quotient by the maximal scaling symmetry $\bZ_{12} \times \bZ_{24}$ we project out 240 of the 243 complex structure moduli and leave the other three untouched. Resolving the orbifold singularities leads to 240 new K\"ahler moduli and the smooth $Y_2$ manifold with Hodge numbers $h^{1,1}=243$ and $h^{2,1}=3$. The interesting feature of this explicit construction is that one can clearly see that the 3 complex structure moduli of $Y_2$ have a moduli space that is a subset of the 243 dimensional complex structure moduli space of $X_2$. This subspace of the complex structure moduli space of $X_2$ is spanned by the three moduli that are invariant under the maximal group of scaling symmetries, the Greene-Plesser (GP) orbifold group. Let us identify them in the defining polynomial of $X_2$. We can write $X_2$ as a hypersurface in $\mathbb{WP}_{1,1,2,8,12}$ (see for example the review \cite{Lusttwo}):
\be\label{eq:defpolX2}
p=\frac{1}{24}(z_1^{24}+z_2^{24}+2 z_3^{12}+8z_4^3+12 z_5^2) - \psi_0 z_1 z_2 z_3 z_4 z_5- \frac16 \psi_1 \ (z_1 z_2 z_3)^6 -\frac{1}{12} \psi_2 \ (z_1 z_2)^{12}\,,
\ee
where $z_i \in \mathbb{WP}_{1,1,2,8,12}$ and the three $\psi_i$ are three of the 243 complex structure moduli.\footnote{In Appendix \ref{app:mirrorsym}, the $\psi_i$ and numerical coefficients together are called $a_i$, with one $a_i$ multiplying each monomial.} The other complex structure moduli correspond to deformations of the polynomial $p$ that we have set to zero. As mentioned above, $X_2$ and therefore $p$ can be quotiented by $G :=\bZ_{12} \times \bZ_{24}$ leading to a singular limit of $Y_2$. From the explicit action of the elements $(g_1, g_2) \in \bZ_{12} \times \bZ_{24}$:
\ba\label{eq:Z24Z12}
&g_1:& (z_1, z_2, z_3, z_4,z_5) \rightarrow (e^{\frac{2\pi i}{12}} z_1, z_2, e^{\frac{2\pi i 11}{12}}z_3, z_4,z_5)\,, \nn\\
&g_2:& (z_1, z_2, z_3, z_4,z_5) \rightarrow (e^{\frac{2\pi i}{24}} z_1, e^{\frac{2\pi i 23}{24}}z_2, z_3, z_4,z_5)\,,
\ea
we see that $p$ is invariant, and therefore the $\psi_i$ correspond to the three complex structure moduli of the mirror manifold $Y_2$ that has Hodge numbers $h^{1,1}=243$, $h^{2,1}=3$. As we have shown in the previous section, these three complex structure moduli are connected to M$_{24}$ and therefore the subset of the complex structure moduli space of $X_2$ that is spanned by the $\psi_i$ is likewise connected to M$_{24}$. Thus we have implicated the K\"ahler moduli space \emph{and} part of the complex structure moduli space of $X_2$ in Mathieu moonshine.

To recap, since we have connected the holomorphic 3-form $\Omega$ of all the $Y_n$ with M$_{24}$, we can now conclude that for the $X_n$ with at least $n=2,4,6,12$ there is also a connection between M$_{24}$ and a subspace of the complex structure moduli space. Similarly, by mirror symmetry this then implies that for the $Y_n$ with at least $n=2,4,6,12$ there is likewise a connection between M$_{24}$ and a subspace of the K\"ahler moduli space.

Note that although the full hypermultiplet moduli spaces of $X_n$ and $Y_n$ are quaternionic K\"ahler, the special slices we discussed in this section (namely, the slice of the complex structure moduli space of $X_n$ and the mirror slice of the K\"ahler moduli space of $Y_n$, and with all RR fields turned off) obey the relations of special K\"ahler geometry. This means for example, that we can calculate period vectors from a prepotential for $X_2$ (and likewise for $X_n$ with $n=4,6,12$). The other polynomial deformations that we have turned off in \eqref{eq:defpolX2} will only appear in the computation of the eight $G$-invariant periods at higher orders, and can be consistently set to zero. This idea was first explored in \cite{Giryavets:2003vd} in the context of flux compactifications.

\section{Mathieu representations in flux compactifications}\label{sec:fluxcomp}
Flux compactifications have been intensively studied during the last fifteen years due their great importance in solving the moduli problem in string compactifications \cite{Grana:2005jc, Douglas:2006es}. The holomorphic 3-form $\Omega$ plays a central role in all flux compactifications on $CY_3$ manifolds that give rise to a four-dimensional $\N=1$ theory due to the Gukov-Vafa-Witten superpotential \cite{Gukov:1999ya}
\be\label{eq:GVW}
W_{GVW} = \int_{CY_3} H_3 \w \Omega\,,
\ee
where $H_3$ denotes the NSNS 3-form flux. In flux compactifications of the heterotic string theory on any of the $Y_n$ (or $X_n$ for $n=2,4,6,12$) we therefore expect the appearance of M$_{24}$ coefficients in the superpotential via the holomorphic 3-form $\Omega$. (As we show in the next section, the superpotential arising in heterotic compactifications on the $X_n$ and $Y_n$ is also connected to M$_{24}$ for $H_3=0$.)

For type II compactifications on a $CY_3$ manifold  one has to do an orientifold projection in order to get a four-dimensional theory with $\N=1$ supersymmetry. For example in type IIA one usually does an orientifold projection that gives rises to $O6$-planes while in type IIB one chooses between either an $O3/O7$ or an $O5/O9$ orientifold projection.\footnote{Depending on the orientifold projection, the four-dimensional $\N=1$ theory might also contain vector multiplets. For type IIB compactifications the resulting holomorphic gauge kinetic function is also related to the holomorphic 3-form $\Omega$ and therefore to M$_{24}$ \cite{Grimm:2004uq, Robbins:2007yv}.} While these orientifold projections can project out some of the complex structure moduli contained in $\Omega$, one generically expects that a connection to M$_{24}$ survives. We work out the details for the most studied class of flux compactifications which is type IIB string theory on a $CY_3$ manifold in the presence of $O3/O7$-planes. In that case the orientifold projection can potentially remove some entries of the period vector but usually all (or the majority) of the entries remain unaffected.

We follow the seminal paper \cite{Giddings:2001yu} that constructs Minkowski vacua in which the complex structure moduli as well as the axio-dilaton are stabilized by fluxes. The reason is that one might wonder whether the appearance of dimensions of M$_{24}$ in the holomorphic 3-form $\Omega$ are due to an actual symmetry of the $Y_n$ and, if that were the case, whether such a symmetry could be a manifest symmetry of the vacua we find in flux compactifications. Due to the large order of M$_{24}$ which is $|$M$_{24}| \approx 2 \times 10^9$ such a symmetry would be very surprising and tremendously interesting. That a sporadic group appears as symmetry group of the internal space used in a string compactifications is of course at the heart of Monstrous moonshine \cite{ConwayNorton}. Monstrous moonshine is essentially explained by the fact that the $\mathbb{Z}_2$ orbifold of $\mathbb{R}^{24}/\Lambda$, where $\Lambda$ is the Leech lattice, has as its symmetry group the Monster group. Compactifying the (left-moving) bosonic string theory on this space leads to a theory with Monster symmetry and the partition function, which is Klein's $J$-function, can therefore be expanded in such a way that the coefficients are (sums of) irreducible representations of the Monster group. Likewise it is clear that the newly discovered mock modular moonshine phenomena involving the Mathieu groups M$_{22}$ and M$_{23}$ \cite{Cheng:2014owa} tell us that superstring compactifications on asymmetric $\mathbb{Z}_2$ orbifolds of $\mathbb{R}^8/\Lambda_{E_8}$, with $\Lambda_{E_8}$ denoting the $E_8$ root lattice, have the symmetry group M$_{22}$ or M$_{23}$. For the case of Mathieu moonshine, however, things are not yet understood and there does not seem to be a direct connection between the Mathieu group M$_{24}$ and the symmetry groups of non-linear sigma models with $\N=(4,4)$ worldsheet symmetry and $K3$ target space \cite{Gaberdiel}. Thus, the fascinating connection between the Gromov-Witten invariants of the $X_n$ and the periods of the holomorphic 3-form $\Omega$ of the $Y_n$ is currently not understood. Nevertheless, it is interesting to understand whether such a symmetry, if it is found to exist, would remain unbroken in flux compactifications. This is what we are explicitly doing for the case of type IIB flux compactifications.

In type IIB flux compactifications on $CY_3$ manifolds we can turn on the NSNS 3-form flux $H_3$ and the RR 3-form flux $F_3$. It is useful to combine these into the complex flux $G_3 = F_3 - \tau H_3$, where $\tau=C_0+i e^{-\phi}$ is the axio-dilaton. We can expand the $G_3$ flux in the basis \eqref{eq:3formbasis} as
\be\label{eq:G3exp}
G_3 = (M^i-\tau \tilde{M}^i) \alpha_i - (N_j -\tau \tilde{N}_j) \beta^j\,,\qquad i=0,1,\ldots,h^{2,1}\,.
\ee
Introducing the flux vectors $f = (N_i,-M^I,-M^0)$ and $h=(\tilde N_i, -\tilde M^I, -\tilde{M}^0)$ where $I=1,2,\ldots,h^{2,1}$, we can write the full flux superpotential as
\be\label{eq:superpot}
W = \int_{CY_3} G_3 \w \Omega = (f-\tau h)\cdot \Pi\,,
\ee
where the period vector $\Pi$ is given in \eqref{eq:pervector}. As we have argued by duality, the instanton numbers (cf. \eqref{eq:matching}) that appear at different powers of $q_i$ in the period vector \eqref{eq:pervector} are related to sums of dimensions of \emph{different} irreducible representations of M$_{24}$. Therefore it seems clear that $\Pi$ does not transform in any well defined way under a potential M$_{24}$ symmetry group. We also notice from equation \eqref{eq:superpot} that $\Pi$ is contracted with a fixed flux vector. This flux vector arises from the expansion of the fluxes in term of 3-forms \eqref{eq:G3exp} and may consist of arbitrary integers, provided they satisfy the tadpole cancellation condition. Since there does not seem to be any M$_{24}$ symmetry acting on the third cohomology class of the $X_n$ or $Y_n$ (cf. \eqref{eq:3formbasis}), the flux vector should be invariant under any potential M$_{24}$ symmetry.

So the lack of a well defined transformation of $\Pi$ together with the contraction with the invariant flux vectors clearly breaks any potential M$_{24}$ symmetry of the $X_n$ or $Y_n$. Thus the resulting flux vacua do therefore not have in any obvious way a large sporadic symmetry group. However, this by no means excludes the exciting possibility that one could define an M$_{24}$ action on the curves that give rise to the Gromov-Witten invariants that seem to be connected the M$_{24}$.

\section{Mathieu representations in Yukawa couplings}\label{sec:Yukawa}
Compactifications of the heterotic string theory on $CY_3$ manifolds give rise to four dimensional $\N=1$ theories with a variety of gauge groups and chiral matter. These compactifications have been studied for decades and have been textbook material for a long time \cite{Polchinski:1998rr}. Here we review a few basic facts and show explicitly how the connection between M$_{24}$ and the Gromov-Witten invariants as well as the holomorphic 3-form $\Omega$ manifests itself in the Yukawa couplings of the four-dimensional theories obtained from compactifying the heterotic string theory on the $X_n$ or $Y_n$.

For compactifications of the heterotic $E_8 \times E_8$ string theory on a $CY_3$ manifold $M$ we have to solve the $H_3$ Bianchi identity which in the absence of NS5-branes reads
\be
dH_3 = \frac{\alpha'}{4} \left[ \text{Tr} \lp R_2 \w R_2 \rp - \text{Tr}_V \lp F_2 \w F_2 \rp \right]\,.
\ee
If we set the gauge connection equal to the spin connection, then this equation is trivially satisfied and all other equations of motion are equally satisfied for $H_3=0$ and constant string coupling. The resulting four dimensional theory preserves $\N=1$ supersymmetry and has a vanishing cosmological constant.
Equating the spin and the gauge connection breaks one of the $E_8$'s to an $E_6$ GUT group and leaves a second unbroken $E_8$. These gauge groups can be further broken by modding out by discrete groups and turning on Wilson lines or by giving expectation values to certain moduli. However, we refrain from doing so to keep the presentation of the connection to M$_{24}$ as transparent as possible. It would be interesting to check whether more involved compactifications on the $X_n$ or $Y_n$ can give rise to semi-realistic models while still preserving the connection to M$_{24}$.

The low energy effective action and the number of chiral multiplets in these compactifications are determined by the topological data of the $CY_3$ manifold $M$. Denoting the Hodge numbers by $h^{p,q}$ one finds $h^{1,1}$ chiral multiplets $\Psi^i$ in the ${\bf 27}$ of $E_6$ and $h^{2,1}$ chiral multiplets $\Phi^\alpha$ in the ${\bf \overline{27}}$ of $E_6$ \cite{Polchinski:1998rr}.\footnote{Here we use different conventions than \cite{Polchinski:1998rr} for ease of presentation.} In addition there are several uncharged chiral multiplets like the $h^{1,1}$ K\"ahler moduli $t^i$, the $h^{2,1}$ complex structure moduli $u^\alpha$ and the axio-dilaton $s$ whose vacuum expectation value controls the tree-level holomorphic gauge kinetic coupling $f^{tree} = s$. The K\"ahler potential for the uncharged K\"ahler and complex structure moduli as well as the axio-dilaton is given by \footnote{We slightly abuse the notation and label the multiplets and the scalar field in the multiplet by the same letter.}
\ba
K_1(t,\bar{t}) &=& -\ln \lp \frac{1}{6}  \int_{M} J \w J \w J \rp = - \ln \lp -\frac{i}{6} \kappa_{ijk}^0 (t^i - \bar{t}^i)(t^j - \bar{t}^j)(t^k - \bar{t}^k)) \rp\,,\\
K_2(u,\bar{u}) &=& \ln \lp i \int_{M} \Omega(u) \w \bar{\Omega}(\bar{u}) \rp \,,\\
K_3(s, \bar{s}) &=& -\ln(s+\bar{s})\,.
\ea
The K\"ahler potential for the matter fields $\Psi^i$ and $\Phi^\alpha$ is
\be
K_{matter} = e^{\frac{K_2-K_1}{3}} \frac{\partial^2 K_1(t,\bar{t})}{\partial t^i \partial \bar{t}^{j}}\Psi^i \bar{\Psi}^{j} + e^{\frac{K_1-K_2}{3}} \frac{\partial^2 K_2(u,\bar{u})}{\partial u^\alpha \partial \bar{u}^{\beta}} \Phi^\alpha \bar{\Phi}^{\beta} \,.
\ee
We see that the holomorphic 3-form $\Omega$ appears in the K\"ahler potential of the four-dimensional theory and therefore M$_{24}$ irreps will appear in the kinetic terms for the $u^\alpha$ and $\Phi^\alpha$ in compactifications on the $Y_n$. Even more interesting is the superpotential. There are non-zero Yukawa couplings for the matter fields that depend on the vacuum expectation values of the uncharged moduli. In particular the superpotential takes the form
\ba
W(t,u,\Psi,\Phi) &=& \frac16 \kappa^0_{ijk}[M] \Psi^i \Psi^j \Psi^k + \frac{1}{6}\frac{\partial^3 F(u)}{\partial u^\alpha\partial u^\beta \partial u^\gamma} \Phi^\alpha \Phi^\beta \Phi^\gamma \nn\\
&=& \frac16 \kappa^0_{ijk}[M] \Psi^i \Psi^j \Psi^k + \frac16 \bar{\kappa}_{\alpha\beta\gamma}[M]\Phi^\alpha \Phi^\beta \Phi^\gamma \,,
\ea
where the gauge indices are contracted with the $E_6$ invariants. We see that the Yukawa couplings for the $\Phi^\alpha$ are derivatives of the prepotential. For compactifications with $M=Y_n$ these are therefore directly related to M$_{24}$. The above K\"ahler and superpotential receive non-perturbative instanton corrections. In particular one expects that the K\"ahler potential $K_1(t,\bar{t})$ and the superpotential for the $\Psi^i$ receive corrections. Due to the invariance under mirror symmetry of these compactifications that preserve $(2,2)$ worldsheet supersymmetry, we expect that these corrections are exactly such that $\kappa^0_{ijk}[M]$ becomes $\kappa_{ijk}[M] $ (cf. equation \eqref{eq:gw}). This means that for compactifications on the $X_n$ the Yukawa couplings for the fields transforming as $\textbf{27}$ are connected to the Mathieu group M$_{24}$ as well, due to the connection between the Gromov-Witten invariants that appear in the instanton-corrected triple intersection numbers and M$_{24}$. As we have argued before at least for $n=2,4,6,12$ there is also a connection between the holomorphic 3-form of the $X_n$ and the Gromov-Witten invariants of the $Y_n$, so at least for these spaces we expect M$_{24}$ to play a role in both Yukawa couplings.

For compactifications of the heterotic string theory on the $Y_n$, we can explicitly calculate the Yukawa couplings in the $STU$ basis up to non-perturbative corrections in $S$, which makes the connection to M$_{24}$ quite transparent. We find the following Yukawa couplings $\bar{\kappa}_{\alpha\beta\gamma}[Y_n] = \partial_\alpha \partial_\beta \partial_\gamma F(S,T,U)$ with $F(S,T,U)$ given in \eqref{eq:prepot} (cf. also \cite{deWit:1995zg, Antoniadis:1995ct})
\ba
\bar{\kappa}_{STU}[Y_n] &=& 1\,,\cr
\bar{\kappa}_{UUU}[Y_n] &=& 2 -2 \sum_{\substack{k>0, l \in \mathbb{Z} \\ k=0,l>0}} c(kl)\ l^3 \lp \frac{1}{1-q_T^k q_U^l} -1\rp= -2  \frac{E_4(q_U) E_4(q_T) E_6(q_T)}{\eta(q_T)^{24} (J(q_U) - J(q_T))}\,,\cr
\bar{\kappa}_{TTT}[Y_n] &=& -2 \sum_{k>0, l \in \mathbb{Z}} c(kl)\ k^3 \lp \frac{1}{1-q_T^k q_U^l} -1\rp = -2 \frac{E_4(q_T) E_4(q_U) E_6(q_U)}{\eta(q_U)^{24} (J(q_T) - J(q_U))}\,,\cr
\bar{\kappa}_{UUT}[Y_n] &=& -2 \sum_{k>0, l \in \mathbb{Z}} c(kl)\ l\ k^2 \lp \frac{1}{1-q_T^k q_U^l} -1\rp\,,\cr
\bar{\kappa}_{UTT}[Y_n] &=& -2 \sum_{k>0, l \in \mathbb{Z}} c(kl)\ l^2 \ k \lp \frac{1}{1-q_T^k q_U^l} -1\rp\,,
\ea
where we used the fact that $\partial_x^3 Li_3(e^x) = \frac{e^x}{1-e^x}$. All other Yukawa couplings vanish perturbatively in $S$. For $\bar{\kappa}_{TTT}[Y_n]$ and $\bar{\kappa}_{UUU}[Y_n]$ a closed form was given in \cite{deWit:1995zg}. There it was also argued that $\bar{\kappa}_{UUT}[Y_n]$ and $\bar{\kappa}_{UTT}[Y_n]$ likewise have a pole for $T=U$ that goes like $(J(q_U) - J(q_T))^{-1}$. However, we did not try to find a closed form for the latter two since the sums make the connection to M$_{24}$ much more transparent. (Recall that the connection between M$_{24}$ and the Yukawa couplings arises due to the relation between the $c(m)$ defined in \eqref{eq:cs} and M$_{24}$; see section \ref{sec:Moonshine}). We thus see that perturbatively in $S$ all non-zero Yukawa couplings, except the trivial $\bar{\kappa}_{STU}[Y_n]$, are linked to M$_{24}$.

From the explicit calculation of the periods that we do in the appendix, we can get the Yukawa couplings to arbitrarily high powers in $q_U$, $q_T$ as well as $q_S$ and we spell them out to a certain order in appendix \ref{app:data}. It is a natural question to ask whether these non-perturbative corrections in $S$ are likewise related to the Mathieu group M$_{24}$. As explained in \cite{Cheng:2013kpa}, based on the recursion relation derived in \cite{Scheidegger, Klemm}, one expects that the answer is yes. Explicitly, on the type IIA side these corrections to the prepotential $F$ that are non-perturbative in $S$ are determined by equations that use as seed the term in $F$ that is perturbative in $S$ and linear in $e^{-2\pi(T-U)}$. This term is nothing but $- 2 E_4 E_6/\eta^{24}$ which is directly related to M$_{24}$ as explained in section \ref{sec:Moonshine}. Thus we see that essentially all terms in the Yukawa couplings are implicated in Mathieu moonshine (albeit in a potentially complicated way).

\section{Conclusion}\label{sec:conclusion}
Mathieu moonshine is an intriguing and not yet understood connection between the elliptic genus of $K3$ and the largest Mathieu group M$_{24}$. In this short paper we extend previous results and explicitly exhibit a link between the periods of certain $CY_3$ manifolds and M$_{24}$. In particular, based on string dualities it was argued in \cite{Cheng:2013kpa} that the Gromov-Witten invariants of the $CY_3$ manifolds $X_n$, that are elliptic fibrations over $\bF_n$, exhibit a connection to M$_{24}$. We extended the checks of this duality that were performed in \cite{Cheng:2013kpa} and argued that this then implies a link between the holomorphic 3-form $\Omega$ of the mirror manifolds $Y_n$ and M$_{24}$. Based on the explicit construction of mirror pairs we have shown that (at least for $n=2,4,6,12$) there is a subspace of the complex structure moduli space for the $X_n$ that is likewise related to M$_{24}$. This then directly implies that a subset of the Gromov-Witten invariants of the $Y_n$ (for at least $n=2,4,6,12$) are also connected to M$_{24}$.

These connections lead to a variety of interesting implications, two of which we discussed in detail. Firstly, flux compactifications on the $CY_3$ manifolds that are implicated in Mathieu moonshine lead to superpotentials with coefficients that are related to the dimensions of representations of M$_{24}$. We noted however that even if these $CY_3$ manifolds have an underlying M$_{24}$ symmetry, then this symmetry should be broken by the Gukov-Vafa-Witten superpotential. Secondly, for simple compactifications of the heterotic $E_8 \times E_8$ string theory on the $CY_3$ manifolds connected to M$_{24}$, we have shown that the Yukawa couplings of the matter fields have an interesting connection to M$_{24}$. In these theories this thus leads to a relation between particle masses and dimensions of representations of the largest Mathieu group M$_{24}$.

It would be interesting to find and study further such connection between physical quantities in four dimensional theories and the Mathieu group M$_{24}$. For example, the action of supersymmetric D6-branes wrapping 3-cycles inside a $CY_3$ manifold involves integrals over the holomorphic 3-form $\Omega$ \cite{Becker:1995kb}. This should lead to a relation between M$_{24}$ and intersecting D6-brane models for compactifications on the $CY_3$ manifolds whose periods are connected to M$_{24}$.

Interestingly, we noticed that $E_4 E_6/ \eta^{24}$ also governs a subclass of Ooguri-Vafa invariants of the three-modulus system composed of the degree-18 $CY_3$ in $\mathbb{WP}_{1, 1, 1, 6, 9}$ and a particular A-brane. See Section 3.2 of \cite{Alim:2009rf} for details of this setup. We noticed that for certain worldsheets wrapping the elliptic fiber of the $CY_3$, these open string analogues of Gromov-Witten invariants were given by exactly $E_4 E_6/ \eta^{24}$.  We computed these invariants to tenth order as a simple check. On the B-model side, the computation of these invariants could be mapped to computations of the periods of a certain $K3$ given as a hypersurface in $\mathbb{WP}_{1, 1, 4, 6}$, much like the $K3$ fiber of the $CY_3$ manifolds studied in this paper! Therefore, it is natural to ask if there is a geometrical explanation for the appearance of this modular form in the periods of these special $K3$s. Of course, the symplectic automorphisms of such $K3$s are strictly subgroups of M$_{23}$, so such an explanation is far from obvious. We may at least be able understand its appearance using restrictions from modularity. While we think such a question is of interest in understanding M$_{24}$'s connection to $K3$ surfaces, it may have further implications for string compactifications as well. In particular, it may suggest that more $CY_3$s (possibly with brane) containing such a $K3$ fiber (or submanifold, up to a change in variables) will have some of its enumerative geometry governed by moonshine. As we discussed in this paper, these invariants manifest in certain quantities in type II and heterotic compactifications.

Relatedly, in \cite{Katz:2014uaa} the authors observe that the dimensions of irreducible representations of $M_{24}$ seem to appear in the stable pair invariants of $K3$ fibered $CY_3$ manifolds. This seems to provide another link between the geometry of $K3$-fibered $CY_3$ manifolds and Mathieu moonshine and it would be very interesting to explore potential connections to our work via the Gromov-Witten/stable pairs correspondence. For example, we do not yet understand how to ``twine" our Gromov-Witten invariants by simple geometric symmetries, and so we cannot compute twining genera to support the connection between moonshine and geometry. The work of \cite{Katz:2014uaa}, however, may suggest  natural geometric twinings, perhaps analogous to the eta-product twinings computed in Mason's moonshine, which would realize an interesting subgroup of M$_{24}$ symmetries acting directly on geometric invariants. This would also be fascinating from the spacetime perspective, as it would translate to an M$_{24}$ action on the algebra of BPS states.

Recently two new moonshine phenomena were discovered in \cite{Cheng:2014owa}. It would be very interesting to understand how they can be connected to explicit string theory compactifications. This should undoubtedly give rise to new interesting physical and mathematical connections involving the Mathieu groups M$_{22}$ and M$_{23}$.

\acknowledgments
We would like to thank N. ~Benjamin, M.~Cheng, X.~Dong, J.~Duncan, S.~Harrison, S.~Kachru, A.~Westphal and D.~Whalen for illuminating discussions on related topics and S.~Kachru for very valuable comments on the manuscript. N.P. is supported by a Stanford Humanities and Sciences Fellowship and T.W. by a Research Fellowship (Grant number WR 166/1-1) of the German Research Foundation (DFG).

\appendix

\section{Brief Review of Mirror Symmetry}\label{app:mirrorsym}

In this section, we will provide a brief review of some of the basic techniques in mirror symmetry that we used in our computations. Our presentation will mostly follow \cite{hosono1995mirror, hosono1995mirror2, hosono1994lectures} and will use the notation of \cite{hosono1994lectures}. For a comprehensive review of mirror symmetry, see the excellent text \cite{hori2003mirror}. For an explicit computation in the one-modulus example of the quintic, we refer the reader to the seminal paper \cite{candelas1991pair}.

Mirror symmetry relates the A- and B-model topological string theories on the mirror manifolds $X_n$, $Y_n$. The A-model is sensitive only to K\"ahler deformations and hence computes the Gromov-Witten invariants on $X_n$; the B-model probes the complex structure moduli space through variations of the Hodge structure. The mirror manifolds are topologically distinct, and their Hodge structures map to one another via a diagonal reflection on their Hodge diamonds. One computes a ``mirror map" $t^i$, which is nothing but a special set of local coordinates, to relate the two theories.

\subsection{Toric data}
In this paper, we focus on closed string mirror symmetry between two $CY_3$ manifolds representable as hypersurfaces in toric varieties. The hypersurfaces are specified by reflexive rational convex polyhedra $(\Delta, \Delta^*)$ and their associated rational fans. The polyhedra will contain the origin, which we denote $\nu_0$. Other integral points in $\Delta$, including vertices, will be denoted $\nu_i$. Given a \textit{reflexive} polyhedron $\Delta$ as a function of the weights, $w_i$, of an ambient weighted projective space $W = \mathbb{WP}_{w_1,w_2,w_3,w_4,w_5}$ one can construct its dual, $\Delta^{*}$, which specifies the topological data of the mirror Calabi-Yau. This is a convenient algorithmic language for finding mirror manifolds which reproduces and extends the Greene-Plesser procedure, which constructs mirrors by orbifolding $X_n$ by some abelian group \cite{greene1990duality}(see also \cite{greene1998new} for a procedure to find mirrors away from the Fermat point). For example, Batyrev \cite{batyrev1994dual} found a simple formula computing the Hodge numbers of the mirror pair in terms of the numbers of integral points on the  faces and interiors of the polyhedra.

If the polyhedron is Gorenstein \footnote{The polyhedron will be Gorenstein if the least common multiple of all the weights $w_i$ divides the degree $d$ of the hypersurface.}, as are the $X_n, n=2, 4, 6, 12$, the dual is simply given by:
\be
\Delta^*(w):= \left\lbrace (x_1, \ldots, x_5) \in \mathbb{R}^5 | \sum_{i=1}^5 w_i x_i =0, x_i \geq -1 \right\rbrace \,.
\ee
In this case, the origin is the only interior point of $\Delta$. Note that the polyhedra satisfy $(\Delta^*)^* = \Delta $.

Normally, we define a hypersurface in weighted projective space as the zero locus of a quasi-homogeneous polynomial $p(z)=0$, which will be nonsingular if it satisfies the transversality conditions. That is, it never fulfills $p(z_p) = dp(z_p)=0$ for any point $z_p$. We can define a toric hypersurface in $W^*$ as the zero locus of the Laurent polynomial $f_{\Delta^*}(a, X) = a_0 - \sum_{i} a_i X^{\nu^*_i}, f_{\Delta^*} \in \mathbb{C}(X_1^{\pm 1}, \ldots, X_4^{\pm 1})$, where $\nu^*_i$ are the integral points in $\Delta^*$ and $a_i$ are complex constants parametrizing the complex structure deformations of the B-model geometry. We have used the notation $X^{\nu_i^*} := \prod_{j=1}^4 X_j^{\nu^*_{i, j}}$. Similar definitions hold for the dual (unstarred) quantities.

For some Calabi-Yau $X$, its complex structure moduli space is encapsulated by lattice points in the polyhedron $\Delta^*$. Each lattice point corresponds to a monomial perturbation. Points in the dual polyhedron correspond to exceptional divisors and therefore encode the K\"ahler moduli space. Mirror symmetry says that if two Calabi-Yaus $X$ and $Y$ are a mirror pair, each realized by a toric hypersurface as described above, then the polyhedra associated to $X$, $(\Delta_X, \Delta^*_X)$, are isomorphic to the polyhedra associated to $Y$, $(\Delta^*_Y, \Delta_Y)$. This exchanges the complex structure and K\"ahler moduli spaces. For simplicity of notation, we have dropped the $X, Y$ subscripts above and in what follows, since we will only care about the pair $(\Delta_X, \Delta^*_Y)$. In this way, we differ slightly from the notation of \cite{hosono1994lectures}, but hope our meaning is clear.

The last important toric quantity to introduce is the Mori cone. There are $5 + h^{2, 1}$ integral points $\nu^*_i$, including the origin $\nu^*_0$ that do not lie in the interior of faces of codimension one. These are the points that we used to construct the Laurent polynomial above. We define a lattice of relations of the form $\sum_i l_i \nu^*_i = 0, l_i \in \mathbb{Z}$. There are $h^{2, 1}(Y_n)(=3$ for our computations on the B-model side) generators of this lattice. Once we find this lattice, we define extended vectors $(l_0^{\alpha}, \left\lbrace l_i^{\alpha} \right\rbrace) := (-\sum_i l_i^{\alpha}, \left\lbrace l_i^{\alpha} \right\rbrace)$. The Mori cone generates the lattice of relations and it will show up in the computation of the periods.

\subsection{Periods and Picard-Fuchs equations}
With topological data in hand, we may now study the B-model on $Y_n$ to extract its holomorphic $(3, 0)$ form $\Omega$ and compute the periods thereof. Following the previous section, this is the mirror manifold associated to $\Delta^*$ so we will explicitly use the $*$ notation to label toric quantities. The period integrals are given by
\be
\Pi(a) = \int_{\gamma_i} \frac{a_0}{f(a, X)} \prod_{j=1}^n \frac{dX_j}{X_j}\,,
\ee
where $f_{\Delta^*}(a, X) = a_0 - \sum_{i} a_i X^{\nu_i^*}$ is the defining polynomial for the hypersurface in terms of complex structure moduli $a_i$ and $X_j$ are inhomogeneous coordinates on $(\mathbb{C^*})^4$ in the ambient projective space. We have again employed the common notation $X^{\nu_i^*} := \prod_j X_j^{\nu^*_{i, j}}$. The number of periods is $dim(H^3) = 2(h^{2, 1}(Y_n) + 1) = 2(h^{1, 1}(X_n) + 1)$, which equals 8 for $X_n$ being an elliptic fibrations over $\mathbb{F}_n$ and $n=0,1,2$.

The periods are solutions to the Picard-Fuchs equations and are readily computable in the large complex structure limit, or around the point of maximal unipotent monodromy. This point will be mapped to the large radius limit of $X_n$ via the mirror map. Given the Mori cone and complex structure moduli, it is convenient to define the variables $u_\alpha := \prod a_i^{l^\alpha_i}, \alpha= 1, \ldots, h^{1, 1}(X_n)$. The large complex structure point is then $u_\alpha=0$. First, one computes the fundamental period directly by choosing the cycle $\Gamma = \left\lbrace (X_1, X_2, X_3, X_4 \in \mathbb{C}^4)\big| |X_i|=1 \right\rbrace$ and computing the integral in the $a_0 \rightarrow \infty$ limit. The result is
\be
w_0(u) = \sum_{n_\alpha}\frac{(- \sum_{\alpha}l_0^{\alpha} n_{\alpha})!}{\prod_{i>0} (l_i^{\alpha} n_{\alpha})!} \prod_{\alpha}u_\alpha^{n_{\alpha}}\,,
\ee
where the sum is such that the integral $n_{\alpha}$ do not let the arguments of the factorials become non-negative.

Now we may set up the GKZ hypergeometric system of partial differential equations which the fundamental period satisfies and a subset of this solution space is the solution space of the Picard-Fuchs (PF) system itself. By examining recursion relations satisfied by the coefficients of the fundamental period, one can find linear differential operators that annihilate the periods:
\be
\left(p_{\beta}(u_{\alpha} \frac{d}{d u_{\alpha}}, u_{\beta} \frac{d}{d u_{\beta}}) - u_{\beta} q_{\beta}(u_{\alpha} \frac{d}{d u_{\alpha}}, u_{\beta} \frac{d}{d u_{\beta}})\right) w(u) = 0\,,
\ee
where $p$ and $q$ are polynomials in the logarithmic derivatives shown. One may then extract the PF system from this GKZ system (sometimes with difficulty, though it is straightforward in our case).

Now, a variation of Hodge structure will change the type of $\Omega(u)$. We can write the cohomology class $H^3(Y_n) = \bigoplus_{p=0}^3 H^{3-p, p}$ by Hodge decomposition, which will vary over the moduli space of complex structures. Indeed, one may think of $H^3(Y_n)$ as the fiber of a vector bundle over the moduli space of complex structures, equipped with a flat connection called the Gauss-Manin connection. One can derive this connection from the PF equations but we will not do so here. For our purposes, we note that we can identify derivatives of $\Omega$ with Hodge filtration spaces and can find linear combinations of derivatives that span the whole filtration. The dimensions of the spaces $(F^3, F^2/F^3, F^1/F^2, F^0/F^1)$ are $(1, h^{2, 1}, h^{2, 1}, 1)$ and integrating the vector obtained from a section of this filtration gives the period vector. Note that the entry corresponding to the 1-dimensional filtration space $F^3$ is, of course, $\Omega(u)$ itself, and the other entries are logarithmic derivatives thereof.

Let's find the vector of periods from the PF equations more concretely, around the point $u=0$. If we apply the method of Frobenius to the PF equations around this point, the result is one power series solution (the fundamental period), and logarithmic solutions, up to a gauge transformation. We analytically continue the fundamental period by swapping the factorials for gamma functions and we add $h^{2, 1}$ new variables $\rho_{\alpha}$ such that $w_0(u, \rho) = \sum c(n + \rho) u^{n + \rho}$. We recover the fundamental period by setting $\rho=0$. In the language of Frobenius, $\rho$ are called the indices, or solutions to the indicial equations, and they turn out to be maximally degenerate and zero at the point of maximal unipotent monodromy.  Turning the crank, we find that the period vector is
\be
\Pi = \begin{pmatrix}
w_0(u) \\
\frac{1}{2 \pi i}\partial_{\rho_i} w_0 |_{\rho=0} \\
\frac{1}{2}\frac{1}{(2 \pi i)^2} \sum \kappa^{0}_{i j k}[X_n] \partial_{\rho_j} \partial_{\rho_k}w_0 |_{\rho=0} \\
\frac{-1}{6}\frac{1}{(2 \pi i)^3} \sum \kappa^{0}_{i j k}[X_n] \partial_{\rho_i} \partial_{\rho_j} \partial_{\rho_k}w_0 |_{\rho=0}
\end{pmatrix}\,.
\ee
Note that the dimensions are $(1, h^{2, 1}, h^{2, 1}, 1)$ as promised. The constants $\kappa^{0}_{ijk}[X_n]$ turn out to be the classical triple intersection numbers of $X_n$ in a particular basis.

\subsection{Flat coordinates and the mirror map}\label{app:flatcoord}

Before we discuss the mirror map, we first introduce the symplectic basis of $H^3(Y_n, \mathbb{Z})$. Since the moduli space of complex structures enjoys the properties of special geometry, this will be the appropriate basis to reexpress the periods in terms of the holomorphic prepotential. As usual, it is
\be\label{eq:3formbasis}
\int_{A^j} \alpha_i = -\int_{B_i} \beta^j = \int_{Y_n} \alpha_i \wedge \beta^j = \delta^j_i\,,\qquad i=0,1,\ldots, h^{2,1}\,.
\ee
In this basis, the periods are written as
\be\label{eq:zF}
z^i = \int_{A^i} \Omega, \quad \mathcal{F}_i(z) = \int_{B_i} \Omega\,.
\ee
The $z^i$ are the special projective coordinates on the moduli space (not to be confused with the coordinates of $W$) and will be identified with $w_i(u)$. Griffiths transversality gives the condition $\int \Omega \wedge \frac{\partial \Omega}{\partial z^i} = 0$, which implies $\mathcal{F}_i = \frac{ \partial \mathcal{F}}{\partial z^i}$, where $\mathcal{F}$ is the holomorphic prepotential. We can go to a physical gauge by dividing by $z^0$ and defining new coordinates $t^i = z^i/z^0$.

In this basis, the triple intersection numbers are $\bar{\kappa}_{ijk} = \int \Omega \wedge \frac{\partial^3}{\partial t^i t^j t^k} \Omega$. Moreover, the period vector becomes

\be\label{eq:periods}
\begin{pmatrix}
1 \\
t^i \\
\frac{\partial}{\partial t^i}\frac{\mathcal{F}}{(z_0)^2} \\
2 \frac{ \mathcal{F}}{(z^0)^2} - t^i \frac{\partial}{\partial t^i}\frac{ \mathcal{F}}{(z^0)^2}
\end{pmatrix}\,.
\ee
The mirror map is given by identifying the new coordinates $t^i$ with the solutions of the PF equations that are linear in logarithms (i.e. the first subspace of dimension $h^{2, 1}$):
\be
t^i(u) = \frac{w_i(u)}{w_0(u)}\,.
\ee

\subsection{Triple intersection numbers and Gromov-Witten invariants}

As discussed in the previous section, the triple intersection numbers $\bar{\kappa}_{ijk}$ are readily computed once we have found the periods. In terms of the prepotential, these are simply rewritten as $\sum_{l=0}^{h^{2, 1}} (z^l \partial_i \partial_j \partial_k \mathcal{F}_l - \mathcal{F}_l \partial_i \partial_j \partial_k z^l)$. We now wish to find the triple intersection numbers on the mirror manifold $X_n$.

If we define $\mathcal{F} = w_0^2 F$, they are:
\be
\kappa_{ijk}[X_n] = \partial_{t^i} \partial_{t^j} \partial_{t^k} F(t) = \frac{1}{w_0(u(t))^2} \frac{\partial u_{\alpha}}{\partial t^i} \frac{\partial u_{\beta}}{\partial t^j} \frac{\partial u_{\gamma}}{\partial t^k}\bar{\kappa}_{\alpha \beta \gamma}[Y_n](u(t))\,.
\ee
If we wish to express the triple intersection numbers in terms of $t^i$, which we know to be the K\"ahler moduli in the limit of large radius, we must invert the mirror map. To do this, we define the variable $q_j = e^{2 \pi i t^j}$. Then we can perform a series inversion $u_i(t)$ fairly laboriously order-by-order. For the simple example of the quintic, this is outlined nicely in \cite{morrison1992picard}. For our three-modulus Hirzebruch surfaces, this is best done with a computer program like Mathematica \cite{instanton}.

We can write these full instanton corrected triple intersection numbers as
\be\label{eq:gw}
\kappa_{ijk}[X_n] = \kappa^{0}_{ijk}[X_n] + \sum_{n_i} \dfrac{N(\left\lbrace n_i \right\rbrace)n_i n_j n_k \prod_l q_l^{n_l}}{1-\prod_{l} q_l^{n_l}}\,,
\ee
where $n_i = \int_C h_i \in \mathbb{Z}$, $h_i \in H^{1, 1}(X_n, \mathbb{Z})$. This expression comes from performing a geometric series coming from multiple coverings of the curve $C$. The integers $N(\left\lbrace n_i \right\rbrace )$ then count the number of (isolated, non-singular) rational curves $C$ of degree $\left\lbrace n_i \right \rbrace$. Hence, these are the integral genus-zero Gromov-Witten invariants. This expression follows from the geometrical definition of the corrected triple intersection numbers, using the fact that $\int_C J = \sum t^i n_i$, where $J$ is the K\"ahler form.

We note that the classical contribution to the triple intersection numbers, $\kappa^0_{ijk}$, are given in a basis corresponding to the variables $u_\alpha$.  It is easy to compute them in the basis of harmonic $(1, 1)$ forms $h_{J}, h_{D_1}, \ldots, h_{D_{h^{1,1}-1}}$, which correspond to the complex structure moduli $a_i$. In the toric language, the computation is described explicitly in \cite{hosono1995mirror}. To compute them in the  basis of divisors (or harmonic forms) corresponding to the $u$ variables, we perform the change of variables $h_J = h_1, \ h_{D_i} = \sum_{\alpha} l^{\alpha}_{i+5} h_{\alpha}$.

Lastly, we note that the prepotential can then be written as
\be
\mathcal{F} = (z_0)^2 \left(\frac{\kappa^{0}_{ijk}[X_n]}{6} t^i t^j t^k + (1/2) a_{ij} t^i t^j + b_i t^i + c/2 +   \frac{1}{(2 \pi i)^3} \sum_{(n_i)} N_{(n_i)} Li_{3}(q^{(n_i)})\right)\,,
\ee
where, up to monodromy transformations, $a_{ij}=0, b_i = \frac{1}{24} \int_{X_n} c_2 \wedge h_i, c= \frac{1}{(2 \pi i)^3} \chi(X_n) \zeta(3)$. Substituting this expression into the period vector makes the dependence of the periods on the Gromov-Witten invariants manifest.

\section{Data for Elliptically Fibered Threefolds}\label{app:data}
Here we present some results of our mirror symmetry computations for elliptic fibrations over $\mathbb{F}_n, n=0, 1, 2$. We list the Mori cone generators, classical topological ring, and the Fourier expansion of the triple intersection numbers $\kappa_{ijk}[X_n]=\bar{\kappa}_{ijk}[Y_n]$ to $5^{th}$ order in the moduli $q_1, q_2, q_3$ that are related to the $STU$ moduli as indicated below. From this expansion, one can easily read off the Gromov-Witten invariants, via \eqref{eq:gw}. 
For all three manifolds, $\chi(X_n)= -480, n=0, 1, 2$.

We also list our $b_i = \frac{1}{24} \int_{X_n} c_2 \wedge h_i$, expressed in the same basis of $h_i$ as \cite{hosono1994lectures}, which we describe in \ref{app:mirrorsym}. Observables like the flux superpotential are, of course, independent of basis choices.

\subsection{$\mathbb{F}_0$}
\begin{align}
l^1&=
\begin{pmatrix}
-6 & 3 & 2 & 1& 0 & 0 & 0 & 0
\end{pmatrix}&
l^2&=
\begin{pmatrix}
0 & 0 & 0 & -2 & 1 & 1 & 0 & 0
\end{pmatrix}&
l^3 &=
\begin{pmatrix}
0 & 0 & 0 & -2 & 0 & 0 & 1 & 1
\end{pmatrix}
\end{align}
\be
q_1 = q_U\,, \quad q_2 = \frac{q_S}{q_U}\,, \quad q_3 = \frac{q_T}{q_U}\,.
\ee
\be
24 b_1 = 92, \quad 24 b_2=24, \quad 24 b_3= 24\,.
\ee
\be
\kappa^0_{111}[X_n] = 8\,, \quad  \kappa^0_{112}[X_n] = 2\,, \quad \kappa^0_{113}[X_n] = 2\,, \quad \kappa^0_{123}[X_n] = 1\,.
\ee
\begin{align}
\kappa_{111}[X_n] &= 8 + 480q_1 + 4320q_1^2 + 13440q_1^3 + 35040q_1^4 + 60480q_1^5 + 480q_1q_2 \\
&+ 2263104q_1^2q_2 + 460581120q_1^3q_2 + 30561073920q_1^4q_2 + 4320q_1^2q_2^2 + 460581120q_1^3q_2^2 \nn\\
&+ 480q_1q_3 + 2263104q_1^2q_3 + 460581120q_1^3q_3 + 30561073920q_1^4q_3 + 1440q_1q_2q_3 \nn\\
&- 1808640q_1^2q_2q_3 + 1390953600q_1^3q_2q_3 + 2400q_1q_2^2q_3  - 3617280q_1^2q_2^2q_3 + 3360q_1q_2^3q_3 \nn\\
&+ 4320q_1^2q_3^2 + 460581120q_1^3q_3^2 + 2400q_1q_2q_3^2 - 3617280q_1^2q_2q_3^2 + 16800q_1q_2^2q_3^2\nn\\
&+ 3360q_1q_2q_3^3 +\ldots\nn\\
\kappa_{112}[X_n] &= 2 + 480q_1q_2 + 1131552q_1^2q_2 + 153527040q_1^3q_2 + 7640268480q_1^4q_2 + 4320q_1^2q_2^2 \\
&+ 307054080q_1^3q_2^2 + 1440q_1q_2q_3 - 904320q_1^2q_2q_3 + 463651200q_1^3q_2q_3 + 4800q_1q_2^2q_3 \nn\\
&- 3617280q_1^2q_2^2q_3 + 10080q_1q_2^3q_3 + 2400q_1q_2q_3^2 - 1808640q_1^2q_2q_3^2 + 33600q_1q_2^2q_3^2 \nn\\
&+ 3360q_1q_2q_3^3+\ldots\nn\\
\kappa_{113}[X_n] &= 2 + 480q_1q_3 + 1131552q_1^2q_3 + 153527040q_1^3q_3 + 7640268480q_1^4q_3  \\
&+ 1440q_1q_2q_3 - 904320q_1^2q_2q_3+ 463651200q_1^3q_2q_3 + 2400q_1q_2^2q_3 - 1808640q_1^2q_2^2q_3  \nn\\
&+ 3360q_1q_2^3q_3+ 4320q_1^2q_3^2 + 307054080q_1^3q_3^2 + 4800q_1q_2q_3^2 - 3617280q_1^2q_2q_3^2 \nn\\
&+ 33600q_1q_2^2q_3^2 + 10080q_1q_2q_3^3+\ldots\nn\\
\kappa_{123}[X_n] &= 1 + 1440q_1q_2q_3 - 452160q_1^2q_2q_3 + 154550400q_1^3q_2q_3 + 4800q_1q_2^2q_3 \\
&+ 10080q_1q_2^3q_3 - 1808640q_1^2q_2^2q_3 + 4800q_1q_2q_3^2 - 1808640q_1^2q_2q_3^2 + 67200q_1q_2^2q_3^2\nn\\
& + 10080q_1q_2q_3^3+\ldots\nn\\
\kappa_{133}[X_n] &= 480q_1q_3 + 565776q_1^2q_3 + 51175680q_1^3q_3 + 1910067120q_1^4q_3 + 1440q_1q_2q_3\\
& - 452160q_1^2q_2q_3 + 154550400q_1^3q_2q_3 + 2400q_1q_2^2q_3 - 904320q_1^2q_2^2q_3 + 3360q_1q_2^3q_3 \nn\\
&+ 4320q_1^2q_3^2 + 204702720q_1^3q_3^2 + 9600q_1q_2q_3^2 - 3617280q_1^2q_2q_3^2 + 67200q_1q_2^2q_3^2\nn\\
& + 30240q_1q_2q_3^3 +\ldots\nn\\
\kappa_{222}[X_n] &= -2q_2 + 480q_1q_2 + 282888q_1^2q_2 + 17058560q_1^3q_2 + 477516780q_1^4q_2 - 2q_2^2  \\
&+ 4320q_1^2q_2^2+ 136468480q_1^3q_2^2 - 2q_2^3 - 2q_2^4 - 2q_2^5 - 4q_2q_3 + 1440q_1q_2q_3 \nn\\
&- 226080q_1^2q_2q_3 + 51516800q_1^3q_2q_3 - 48q_2^2q_3 + 19200q_1q_2^2q_3 - 3617280q_1^2q_2^2q_3 \nn\\
&- 216q_2^3q_3 + 90720q_1q_2^3q_3  - 640q_2^4q_3- 6q_2q_3^2 + 2400q_1q_2q_3^2 - 452160q_1^2q_2q_3^2 - 260q_2^2q_3^2   \nn\\
&+ 134400q_1q_2^2q_3^2 - 2970q_2^3q_3^2- 8q_2q_3^3 + 3360q_1q_2q_3^3 - 880q_2^2q_3^3 - 10q_2q_3^4+\ldots\nn\\
\kappa_{223}[X_n] &=  -4q_2q_3 + 1440q_1q_2q_3 - 226080q_1^2q_2q_3 + 51516800q_1^3q_2q_3 - 24q_2^2q_3 \\
&+ 9600q_1q_2^2q_3 - 1808640q_1^2q_2^2q_3 - 72q_2^3q_3 + 30240q_1q_2^3q_3 - 160q_2^4q_3 - 12q_2q_3^2  \nn\\
&+ 4800q_1q_2q_3^2- 904320q_1^2q_2q_3^2 - 260q_2^2q_3^2 + 134400q_1q_2^2q_3^2 - 1980q_2^3q_3^2 - 24q_2q_3^3  \nn\\
&+ 10080q_1q_2q_3^3- 1320q_2^2q_3^3 - 40q_2q_3^4+\ldots\nn\\
\kappa_{233}[X_n] &= -4q_2q_3 + 1440q_1q_2q_3 - 226080q_1^2q_2q_3 + 51516800q_1^3q_2q_3 - 12q_2^2q_3 \\
&+ 4800q_1q_2^2q_3 - 904320q_1^2q_2^2q_3 - 24q_2^3q_3 + 10080q_1q_2^3q_3 - 40q_2^4q_3 - 24q_2q_3^2  \nn\\
&+ 9600q_1q_2q_3^2- 1808640q_1^2q_2q_3^2- 260q_2^2q_3^2 + 134400q_1q_2^2q_3^2 - 1320q_2^3q_3^2 - 72q_2q_3^3 \nn\\
&+ 30240q_1q_2q_3^3 - 1980q_2^2q_3^3 - 160q_2q_3^4 +\ldots\nn\\
\kappa_{333}[X_n] &= -2q_3 + 480q_1q_3 + 282888q_1^2q_3 + 17058560q_1^3q_3 + 477516780q_1^4q_3 - 4q_2q_3 \\
&+ 1440q_1q_2q_3- 226080q_1^2q_2q_3 + 51516800q_1^3q_2q_3 - 6q_2^2q_3 + 2400q_1q_2^2q_3 \nn\\
&- 452160q_1^2q_2^2q_3 - 8q_2^3q_3 + 3360q_1q_2^3q_3 - 10q_2^4q_3- 2q_3^2 + 4320q_1^2q_3^2 + 136468480q_1^3q_3^2 \nn\\
&- 48q_2q_3^2   + 19200q_1q_2q_3^2 - 3617280q_1^2q_2q_3^2- 260q_2^2q_3^2 + 134400q_1q_2^2q_3^2- 880q_2^3q_3^2 \nn\\
&- 2q_3^3 - 216q_2q_3^3 + 90720q_1q_2q_3^3 - 2970q_2^2q_3^3 - 2q_3^4 - 640q_2q_3^4 - 2q_3^5 +\ldots\nn
\end{align}

\subsection{$\mathbb{F}_1$}
\begin{align}
l^1&=
\begin{pmatrix}
-6 & 3 & 2 & 1& 0 & 0 & 0 & 0
\end{pmatrix}&
l^2&=
\begin{pmatrix}
0 & 0 & 0 & -2 & 1 & 1 & 0 & 0
\end{pmatrix}&
l^3 &=
\begin{pmatrix}
0 & 0 & 0 & -1 & 0 & -1 & 1 & 1
\end{pmatrix}
\end{align}
\be
q_1 = q_U\,, \quad q_2 = \frac{q_T}{q_U}\,, \quad q_3 = \frac{q_S}{(q_U q_T)^{\frac12}} \,.
\ee
\be
24 b_1 = 92, \quad 24 b_2= 36, \quad 24 b_3= 24
\ee
\be
\kappa^0_{111}[X_n] = 8\,, \quad  \kappa^0_{112}[X_n] = 3\,, \quad \kappa^0_{122}[X_n] = 1\,,\quad \kappa^0_{113}[X_n] = 2\,, \quad \kappa^0_{123}[X_n] = 1\,.
\ee
\begin{align}
\kappa_{111}[X_n] &= 8 + 480q_1 + 4320q_1^2 + 13440q_1^3 + 35040q_1^4 + 60480q_1^5 + 480q_1q_2 \\
&+ 2263104q_1^2q_2 + 460581120q_1^3q_2 + 30561073920q_1^4q_2 + 4320q_1^2q_2^2 + 460581120q_1^3q_2^2 \nn\\
&+ 252q_1q_3 + 41040q_1^2q_3 + 1478520q_1^3q_3 + 26873280q_1^4q_3 - 960q_1q_2q_3 \nn\\
&+ 945360q_1^2q_2q_3  + 5029579008q_1^3q_2q_3 - 1920q_1q_2^2q_3+ 2712960q_1^2q_2^2q_3 - 2880q_1q_2^3q_3 \nn\\
&- 73764q_1^2q_3^2- 18191520q_1^3q_3^2 - 82080q_1^2q_2q_3^2 + 2400q_1q_2^2q_3^2+\ldots\nn\\
\kappa_{112}[X_n] &= 3 + 480q_1q_2 + 1131552q_1^2q_2 + 153527040q_1^3q_2 + 7640268480q_1^4q_2\\
&+ 4320q_1^2q_2^2  + 307054080q_1^3q_2^2 - 960q_1q_2q_3 + 472680q_1^2q_2q_3 + 1676526336q_1^3q_2q_3 \nn\\
&- 3840q_1q_2^2q_3 + 2712960q_1^2q_2^2q_3  - 8640q_1q_2^3q_3 - 41040q_1^2q_2q_3^2 + 4800q_1q_2^2q_3^2+\ldots\nn\\
\kappa_{113}[X_n] &= 2 + 252q_1q_3 + 20520q_1^2q_3 + 492840q_1^3q_3 + 6718320q_1^4q_3  - 960q_1q_2q_3 \\
&+ 472680q_1^2q_2q_3 + 1676526336q_1^3q_2q_3 - 1920q_1q_2^2q_3 + 1356480q_1^2q_2^2q_3 - 2880q_1q_2^3q_3 \nn\\
&-73764q_1^2q_3^2  - 12127680q_1^3q_3^2 - 82080q_1^2q_2q_3^2 + 4800q_1q_2^2q_3^2+\ldots\nn\\
\kappa_{122}[X_n] &= 1 + 480q_1q_2 + 565776q_1^2q_2 + 51175680q_1^3q_2 + 1910067120q_1^4q_2 \\
&+ 4320q_1^2q_2^2 + 204702720q_1^3q_2^2  - 960q_1q_2q_3 + 236340q_1^2q_2q_3 + 558842112q_1^3q_2q_3 \nn \\
&- 7680q_1q_2^2q_3 + 2712960q_1^2q_2^2q_3 - 25920q_1q_2^3q_3 - 20520q_1^2q_2q_3^2 + 9600q_1q_2^2q_3^2+\ldots\nn\\
\kappa_{123}[X_n] &= 1 - 960q_1q_2q_3 + 236340q_1^2q_2q_3 + 558842112q_1^3q_2q_3 - 3840q_1q_2^2q_3 \\
&+ 1356480q_1^2q_2^2q_3 - 8640q_1q_2^3q_3 - 41040q_1^2q_2q_3^2 + 9600q_1q_2^2q_3^2+\ldots\nn\\
\kappa_{133}[X_n] &= 252q_1q_3 + 10260q_1^2q_3 + 164280q_1^3q_3 + 1679580q_1^4q_3 - 960q_1q_2q_3 \\
&+ 236340q_1^2q_2q_3 + 558842112q_1^3q_2q_3 - 1920q_1q_2^2q_3 + 678240q_1^2q_2^2q_3 - 2880q_1q_2^3q_3\nn\\
& - 73764q_1^2q_3^2 - 8085120q_1^3q_3^2 - 82080q_1^2q_2q_3^2 + 9600q_1q_2^2q_3^2 +\ldots\nn\\
\kappa_{222}[X_n] &= -2q_2 + 480q_1q_2 + 282888q_1^2q_2 + 17058560q_1^3q_2 + 477516780q_1^4q_2 - 2q_2^2 \\
&+ 4320q_1^2q_2^2 + 136468480q_1^3q_2^2 - 2q_2^3 - 2q_2^4 - 2q_2^5 + 3q_2q_3 - 960q_1q_2q_3 + 118170q_1^2q_2q_3 \nn\\
&+ 186280704q_1^3q_2q_3 + 40q_2^2q_3 - 15360q_1q_2^2q_3 + 2712960q_1^2q_2^2q_3  + 189q_2^3q_3 \nn\\
&- 77760q_1q_2^3q_3 + 576q_2^4q_3 - 10260q_1^2q_2q_3^2- 45q_2^2q_3^2 + 19200q_1q_2^2q_3^2 - 864q_2^3q_3^2 +\ldots\nn\\
\kappa_{223}[X_n] &= 3q_2q_3 - 960q_1q_2q_3 + 118170q_1^2q_2q_3 + 186280704q_1^3q_2q_3 + 20q_2^2q_3\\
& - 7680q_1q_2^2q_3 + 1356480q_1^2q_2^2q_3 + 63q_2^3q_3 - 25920q_1q_2^3q_3 + 144q_2^4q_3 - 20520q_1^2q_2q_3^2\nn\\
& - 45q_2^2q_3^2 + 19200q_1q_2^2q_3^2 - 576q_2^3q_3^2 +\ldots\nn\\
\kappa_{233}[X_n] &= 3q_2q_3 - 960q_1q_2q_3 + 118170q_1^2q_2q_3 + 186280704q_1^3q_2q_3 + 10q_2^2q_3\\
& - 3840q_1q_2^2q_3 + 678240q_1^2q_2^2q_3 + 21q_2^3q_3 - 8640q_1q_2^3q_3 + 36q_2^4q_3 - 41040q_1^2q_2q_3^2\nn\\
& - 45q_2^2q_3^2 + 19200q_1q_2^2q_3^2 - 384q_2^3q_3^2+\ldots\nn\\
\kappa_{333}[X_n] &= q_3 + 252q_1q_3 + 5130q_1^2q_3 + 54760q_1^3q_3 + 419895q_1^4q_3 + 3q_2q_3 - 960q_1q_2q_3 \\
&+ 118170q_1^2q_2q_3 + 186280704q_1^3q_2q_3 + 5q_2^2q_3 - 1920q_1q_2^2q_3 + 339120q_1^2q_2^2q_3 + 7q_2^3q_3 \nn\\
&- 2880q_1q_2^3q_3 + 9q_2^4q_3 + q_3^2 - 73764q_1^2q_3^2 - 5390080q_1^3q_3^2 - 82080q_1^2q_2q_3^2 - 45q_2^2q_3^2  \nn\\
&+ 19200q_1q_2^2q_3^2- 256q_2^3q_3^2 + q_3^3 + q_3^4 + q_3^5+\ldots\nn
\end{align}

\subsection{$\mathbb{F}_2$}
\begin{align}
l^1&=
\begin{pmatrix}
-6 & 3 & 2 & 1& 0 & 0 & 0 & 0
\end{pmatrix}&
l^2&=
\begin{pmatrix}
0 & 0 & 0 & -2 & 1 & 1 & 0 & 0
\end{pmatrix}&
l^3 &=
\begin{pmatrix}
0 & 0 & 0 & 0 & 0 & -2 & 1 & 1
\end{pmatrix}
\end{align}
\be
q_1 = q_U\,, \quad q_2 = \frac{q_T}{q_U}\,, \quad q_3 = \frac{q_S}{q_T}\,.
\ee
\be
24 b_1 = 92, \quad 24 b_2=48, \quad 24 b_3= 24
\ee
\be
\kappa^0_{111}[X_n] = 8\,, \quad  \kappa^0_{112}[X_n] = 4\,, \quad \kappa^0_{122}[X_n] = 2\,,\quad \kappa^0_{113}[X_n] = 2\,, \quad \kappa^0_{123}[X_n] = 1\,.
\ee
\begin{align}
\kappa_{111}[X_n] &= 8 + 480q_1 + 4320q_1^2 + 13440q_1^3 + 35040q_1^4 + 60480q_1^5 + 480q_1q_2 \\
&+ 2263104q_1^2q_2 + 460581120q_1^3q_2 + 30561073920q_1^4q_2 + 4320q_1^2q_2^2 + 460581120q_1^3q_2^2\nn \\
&+ 480q_1q_2q_3 + 2263104q_1^2q_2q_3 + 460581120q_1^3q_2q_3 + 1440q_1q_2^2q_3 - 1808640q_1^2q_2^2q_3 \nn\\
&+ 2400q_1q_2^3q_3 +\ldots\nn\\
\kappa_{112}[X_n] &= 4 + 480q_1q_2 + 1131552q_1^2q_2 + 153527040q_1^3q_2 + 7640268480q_1^4q_2 \\
& + 153527040q_1^3q_2q_3 + 4320q_1^2q_2^2 + 307054080q_1^3q_2^2 + 480q_1q_2q_3 + 1131552q_1^2q_2q_3 \nn\\
& + 2880q_1q_2^2q_3 - 1808640q_1^2q_2^2q_3 + 7200q_1q_2^3q_3+\ldots\nn \\
\kappa_{113}[X_n] &= 2 + 480q_1q_2q_3 + 1131552q_1^2q_2q_3 + 153527040q_1^3q_2q_3 + 1440q_1q_2^2q_3 \\
&- 904320q_1^2q_2^2q_3 + 2400q_1q_2^3q_3+\ldots\nn\\
\kappa_{122}[X_n] &= 2 + 480q_1q_2 + 565776q_1^2q_2 + 51175680q_1^3q_2 + 1910067120q_1^4q_2\\
&  + 4320q_1^2q_2^2 + 204702720q_1^3q_2^2 + 480q_1q_2q_3 + 565776q_1^2q_2q_3 + 51175680q_1^3q_2q_3 \nn \\
&+ 5760q_1q_2^2q_3 - 1808640q_1^2q_2^2q_3 + 21600q_1q_2^3q_3+\ldots\nn\\
\kappa_{123}[X_n] &= 1 + 480q_1q_2q_3 + 565776q_1^2q_2q_3 + 51175680q_1^3q_2q_3 + 2880q_1q_2^2q_3 \\
&- 904320q_1^2q_2^2q_3 + 7200q_1q_2^3q_3+\ldots\nn\\
\kappa_{133}[X_n] &= 480q_1q_2q_3 + 565776q_1^2q_2q_3 + 51175680q_1^3q_2q_3 + 1440q_1q_2^2q_3 \\
&- 452160q_1^2q_2^2q_3 + 2400q_1q_2^3q_3+\ldots\nn\\
\kappa_{222}[X_n] &= -2q_2 + 480q_1q_2 + 282888q_1^2q_2 + 17058560q_1^3q_2 + 477516780q_1^4q_2 - 2q_2^2\\
& + 4320q_1^2q_2^2 + 136468480q_1^3q_2^2 - 2q_2^3 - 2q_2^4 - 2q_2^5 - 2q_2q_3 + 480q_1q_2q_3 + 282888q_1^2q_2q_3 \nn\\
&+ 17058560q_1^3q_2q_3 - 32q_2^2q_3 + 11520q_1q_2^2q_3 - 1808640q_1^2q_2^2q_3 - 162q_2^3q_3 \nn\\
&+ 64800q_1q_2^3q_3 - 512q_2^4q_3 - 2q_2^2q_3^2 - 162q_2^3q_3^2+\ldots\nn\\
\kappa_{223}[X_n] &= -2q_2q_3 + 480q_1q_2q_3 + 282888q_1^2q_2q_3 + 17058560q_1^3q_2q_3 + 5760q_1q_2^2q_3 \\
&- 16q_2^2q_3  - 904320q_1^2q_2^2q_3  - 54q_2^3q_3 + 21600q_1q_2^3q_3 - 128q_2^4q_3 - 2q_2^2q_3^2 - 108q_2^3q_3^2+\ldots\nn\\
\kappa_{233}[X_n] &= -2q_2q_3 + 480q_1q_2q_3 + 282888q_1^2q_2q_3 + 17058560q_1^3q_2q_3 - 8q_2^2q_3 \\
&+ 2880q_1q_2^2q_3 - 452160q_1^2q_2^2q_3 - 18q_2^3q_3 + 7200q_1q_2^3q_3 - 32q_2^4q_3 - 2q_2^2q_3^2 - 72q_2^3q_3^2+\ldots\nn\\
\kappa_{333}[X_n] &= -2q_2q_3 + 480q_1q_2q_3 + 282888q_1^2q_2q_3 + 17058560q_1^3q_2q_3 - 4q_2^2q_3 \\
& + 1440q_1q_2^2q_3 - 226080q_1^2q_2^2q_3 - 6q_2^3q_3 + 2400q_1q_2^3q_3 - 8q_2^4q_3 - 2q_2^2q_3^2 - 48q_2^3q_3^2+\ldots\nn
\end{align}

\bibliographystyle{JHEP}
\bibliography{refs}

\providecommand{\href}[2]{#2}\begingroup\raggedright\begin{thebibliography}{10}

\bibitem{EOT}
T.~Eguchi, H.~Ooguri, and Y.~Tachikawa, {\it {Notes on the K3 Surface and the
  Mathieu group $M_{24}$}},  {\em Exper.Math.} {\bf 20} (2011) 91--96,
  [\href{http://xxx.lanl.gov/abs/1004.0956}{{\tt arXiv:1004.0956}}].

\bibitem{Miranda}
M.~C. Cheng, {\it {K3 Surfaces, N=4 Dyons, and the Mathieu Group M24}},  {\em
  Commun.Num.Theor.Phys.} {\bf 4} (2010) 623--658,
  [\href{http://xxx.lanl.gov/abs/1005.5415}{{\tt arXiv:1005.5415}}].

\bibitem{Gaberdiel:2010ch}
M.~R. Gaberdiel, S.~Hohenegger, and R.~Volpato, {\it {Mathieu twining
  characters for K3}},  {\em JHEP} {\bf 1009} (2010) 058,
  [\href{http://xxx.lanl.gov/abs/1006.0221}{{\tt arXiv:1006.0221}}].

\bibitem{Gaberdiel:2010ca}
M.~R. Gaberdiel, S.~Hohenegger, and R.~Volpato, {\it {Mathieu Moonshine in the
  elliptic genus of K3}},  {\em JHEP} {\bf 1010} (2010) 062,
  [\href{http://xxx.lanl.gov/abs/1008.3778}{{\tt arXiv:1008.3778}}].

\bibitem{Eguchi:2010fg}
T.~Eguchi and K.~Hikami, {\it {Note on Twisted Elliptic Genus of K3 Surface}},
  {\em Phys.Lett.} {\bf B694} (2011) 446--455,
  [\href{http://xxx.lanl.gov/abs/1008.4924}{{\tt arXiv:1008.4924}}].

\bibitem{Cheng:2012tq}
M.~C. Cheng, J.~F. Duncan, and J.~A. Harvey, {\it {Umbral Moonshine}},
  \href{http://xxx.lanl.gov/abs/1204.2779}{{\tt arXiv:1204.2779}}.

\bibitem{Cheng:2013wca}
M.~C.~N. Cheng, J.~F.~R. Duncan, and J.~A. Harvey, {\it {Umbral Moonshine and
  the Niemeier Lattices}},  \href{http://xxx.lanl.gov/abs/1307.5793}{{\tt
  arXiv:1307.5793}}.

\bibitem{Cheng:2014zpa}
M.~C.~N. Cheng and S.~Harrison, {\it {Umbral Moonshine and K3 Surfaces}},
  \href{http://xxx.lanl.gov/abs/1406.0619}{{\tt arXiv:1406.0619}}.

\bibitem{Persson:2013xpa}
D.~Persson and R.~Volpato, {\it {Second Quantized Mathieu Moonshine}},
  \href{http://xxx.lanl.gov/abs/1312.0622}{{\tt arXiv:1312.0622}}.

\bibitem{Gannonproof}
T.~Gannon, {\it {Much ado about Mathieu}},
  \href{http://xxx.lanl.gov/abs/1211.5531}{{\tt arXiv:1211.5531}}.

\bibitem{Gaberdiel}
M.~R. Gaberdiel, S.~Hohenegger, and R.~Volpato, {\it {Symmetries of K3 sigma
  models}},  {\em Commun.Num.Theor.Phys.} {\bf 6} (2012) 1--50,
  [\href{http://xxx.lanl.gov/abs/1106.4315}{{\tt arXiv:1106.4315}}].

\bibitem{Taormina:2011rr}
A.~Taormina and K.~Wendland, {\it {The overarching finite symmetry group of
  Kummer surfaces in the Mathieu group $M_{24}$}},  {\em JHEP} {\bf 1308}
  (2013) 125, [\href{http://xxx.lanl.gov/abs/1107.3834}{{\tt
  arXiv:1107.3834}}].

\bibitem{Taormina:2013jza}
A.~Taormina and K.~Wendland, {\it {Symmetry-surfing the moduli space of Kummer
  K3s}},  \href{http://xxx.lanl.gov/abs/1303.2931}{{\tt arXiv:1303.2931}}.

\bibitem{Cheng:2013kpa}
M.~C. Cheng, X.~Dong, J.~Duncan, J.~Harvey, S.~Kachru, and T.~Wrase, {\it
  {Mathieu Moonshine and N=2 String Compactifications}},  {\em JHEP} {\bf 1309}
  (2013) 030, [\href{http://xxx.lanl.gov/abs/1306.4981}{{\tt
  arXiv:1306.4981}}].

\bibitem{Harrison:2013bya}
S.~Harrison, S.~Kachru, and N.~M. Paquette, {\it {Twining Genera of (0,4)
  Supersymmetric Sigma Models on K3}},  {\em JHEP} {\bf 1404} (2014) 048,
  [\href{http://xxx.lanl.gov/abs/1309.0510}{{\tt arXiv:1309.0510}}].

\bibitem{Hohenegger:2011us}
S.~Hohenegger and S.~Stieberger, {\it {BPS Saturated String Amplitudes: K3
  Elliptic Genus and Igusa Cusp Form}},  {\em Nucl.Phys.} {\bf B856} (2012)
  413--448, [\href{http://xxx.lanl.gov/abs/1108.0323}{{\tt arXiv:1108.0323}}].

\bibitem{Harvey:2013mda}
J.~A. Harvey and S.~Murthy, {\it {Moonshine in Fivebrane Spacetimes}},  {\em
  JHEP} {\bf 1401} (2014) 146, [\href{http://xxx.lanl.gov/abs/1307.7717}{{\tt
  arXiv:1307.7717}}].

\bibitem{Wrase:2014fja}
T.~Wrase, {\it {Mathieu moonshine in four dimensional $\mathcal{N}=1$
  theories}},  {\em JHEP} {\bf 1404} (2014) 069,
  [\href{http://xxx.lanl.gov/abs/1402.2973}{{\tt arXiv:1402.2973}}].

\bibitem{greene1990duality}
B.~R. Greene and M.~R. Plesser, {\it Duality in calabi-yau moduli space},  {\em
  Nuclear Physics B} {\bf 338} (1990), no.~1 15--37.

\bibitem{Eguchi:1988vra}
T.~Eguchi, H.~Ooguri, A.~Taormina, and S.-K. Yang, {\it {Superconformal
  Algebras and String Compactification on Manifolds with SU(N) Holonomy}},
  {\em Nucl.Phys.} {\bf B315} (1989) 193.

\bibitem{Lusttwo}
D.~L{\"u}st, {\it {String vacua with N=2 supersymmetry in four-dimensions}},
  \href{http://xxx.lanl.gov/abs/hep-th/9803072}{{\tt hep-th/9803072}}.

\bibitem{Klemm}
A.~Klemm, J.~Manschot, and T.~Wotschke, {\it {Quantum geometry of elliptic
  Calabi-Yau manifolds}},  \href{http://xxx.lanl.gov/abs/1205.1795}{{\tt
  arXiv:1205.1795}}.

\bibitem{Scheidegger}
M.~Alim and E.~Scheidegger, {\it {Topological Strings on Elliptic Fibrations}},
   \href{http://xxx.lanl.gov/abs/1205.1784}{{\tt arXiv:1205.1784}}.

\bibitem{hosono1995mirror}
S.~Hosono, A.~Klemm, S.~Thiesen, and S.-T. Yau, {\it Mirror symmetry, mirror
  map and applications to calabi-yau hypersurfaces},  {\em Communications in
  Mathematical Physics} {\bf 167} (1995), no.~2 301--350.

\bibitem{Gukov:1999ya}
S.~Gukov, C.~Vafa, and E.~Witten, {\it {CFT's from Calabi-Yau four folds}},
  {\em Nucl.Phys.} {\bf B584} (2000) 69--108,
  [\href{http://xxx.lanl.gov/abs/hep-th/9906070}{{\tt hep-th/9906070}}].

\bibitem{Giryavets:2003vd}
A.~Giryavets, S.~Kachru, P.~K. Tripathy, and S.~P. Trivedi, {\it {Flux
  compactifications on Calabi-Yau threefolds}},  {\em JHEP} {\bf 0404} (2004)
  003, [\href{http://xxx.lanl.gov/abs/hep-th/0312104}{{\tt hep-th/0312104}}].

\bibitem{Grana:2005jc}
M.~Grana, {\it {Flux compactifications in string theory: A Comprehensive
  review}},  {\em Phys.Rept.} {\bf 423} (2006) 91--158,
  [\href{http://xxx.lanl.gov/abs/hep-th/0509003}{{\tt hep-th/0509003}}].

\bibitem{Douglas:2006es}
M.~R. Douglas and S.~Kachru, {\it {Flux compactification}},  {\em
  Rev.Mod.Phys.} {\bf 79} (2007) 733--796,
  [\href{http://xxx.lanl.gov/abs/hep-th/0610102}{{\tt hep-th/0610102}}].

\bibitem{Grimm:2004uq}
T.~W. Grimm and J.~Louis, {\it {The Effective action of N = 1 Calabi-Yau
  orientifolds}},  {\em Nucl.Phys.} {\bf B699} (2004) 387--426,
  [\href{http://xxx.lanl.gov/abs/hep-th/0403067}{{\tt hep-th/0403067}}].

\bibitem{Robbins:2007yv}
D.~Robbins and T.~Wrase, {\it {D-terms from generalized NS-NS fluxes in type
  II}},  {\em JHEP} {\bf 0712} (2007) 058,
  [\href{http://xxx.lanl.gov/abs/0709.2186}{{\tt arXiv:0709.2186}}].

\bibitem{Giddings:2001yu}
S.~B. Giddings, S.~Kachru, and J.~Polchinski, {\it {Hierarchies from fluxes in
  string compactifications}},  {\em Phys.Rev.} {\bf D66} (2002) 106006,
  [\href{http://xxx.lanl.gov/abs/hep-th/0105097}{{\tt hep-th/0105097}}].

\bibitem{ConwayNorton}
J.~H. Conway and S.~P. Norton, {\it {Monstrous Moonshine}},  {\em Bull. London
  Math. Soc.} {\bf 11} (1979) 308--339.

\bibitem{Cheng:2014owa}
M.~C.~N. Cheng, X.~Dong, J.~F.~R. Duncan, S.~Harrison, S.~Kachru, and T.~Wrase,
  {\it {Mock Modular Mathieu Moonshine Modules}},
  \href{http://xxx.lanl.gov/abs/1406.5502}{{\tt arXiv:1406.5502}}.

\bibitem{Polchinski:1998rr}
J.~Polchinski, {\it {String theory. Vol. 2: Superstring theory and beyond}}, .

\bibitem{deWit:1995zg}
B.~de~Wit, V.~Kaplunovsky, J.~Louis, and D.~L{\"u}st, {\it {Perturbative
  couplings of vector multiplets in N=2 heterotic string vacua}},  {\em
  Nucl.Phys.} {\bf B451} (1995) 53--95,
  [\href{http://xxx.lanl.gov/abs/hep-th/9504006}{{\tt hep-th/9504006}}].

\bibitem{Antoniadis:1995ct}
I.~Antoniadis, S.~Ferrara, E.~Gava, K.~Narain, and T.~Taylor, {\it
  {Perturbative prepotential and monodromies in N=2 heterotic superstring}},
  {\em Nucl.Phys.} {\bf B447} (1995) 35--61,
  [\href{http://xxx.lanl.gov/abs/hep-th/9504034}{{\tt hep-th/9504034}}].

\bibitem{Becker:1995kb}
K.~Becker, M.~Becker, and A.~Strominger, {\it {Five-branes, membranes and
  nonperturbative string theory}},  {\em Nucl.Phys.} {\bf B456} (1995)
  130--152, [\href{http://xxx.lanl.gov/abs/hep-th/9507158}{{\tt
  hep-th/9507158}}].

\bibitem{Alim:2009rf}
M.~Alim, M.~Hecht, P.~Mayr, and A.~Mertens, {\it {Mirror Symmetry for Toric
  Branes on Compact Hypersurfaces}},  {\em JHEP} {\bf 0909} (2009) 126.

\bibitem{Katz:2014uaa}
S.~Katz, A.~Klemm, and R.~Pandharipande, {\it {On the motivic stable pairs
  invariants of K3 surfaces}},  \href{http://xxx.lanl.gov/abs/1407.3181}{{\tt
  arXiv:1407.3181}}.

\bibitem{hosono1995mirror2}
S.~Hosono, A.~Klemm, S.~Theisen, and S.-T. Yau, {\it Mirror symmetry, mirror
  map and applications to complete intersection calabi-yau spaces},  {\em
  Nuclear Physics B} {\bf 433} (1995), no.~3 501--552.

\bibitem{hosono1994lectures}
S.~Hosono, A.~Klemm, and S.~Theisen, {\it Lectures on mirror symmetry},  in
  {\em Integrable models and strings}, pp.~235--280.
\newblock Springer, 1994.

\bibitem{hori2003mirror}
K.~Hori, {\em Mirror symmetry}, vol.~1.
\newblock American Mathematical Soc., 2003.

\bibitem{candelas1991pair}
P.~Candelas, X.~C. De~La~Ossa, P.~S. Green, and L.~Parkes, {\it A pair of
  calabi-yau manifolds as an exactly soluble superconformal theory},  {\em
  Nuclear Physics B} {\bf 359} (1991), no.~1 21--74.

\bibitem{greene1998new}
B.~R. Greene, M.~Plesser, and S.~Roan, {\it New constructions of mirror
  manifolds: Probing moduli space far from fermat points},  {\em AMS/IP Studies
  in Advanced Mathematics} {\bf 9} (1998) 347--390.

\bibitem{batyrev1994dual}
V.~V. Batyrev, {\it Dual polyhedra and mirror symmetry for calabi--yau
  hypersurfaces in toric varieties},  in {\em J. Alg. Geom}, Citeseer, 1994.

\bibitem{morrison1992picard}
D.~R. Morrison, {\it {Picard-Fuchs equations and mirror maps for
  hypersurfaces}},  \href{http://xxx.lanl.gov/abs/alg-geom/9202026}{{\tt
  alg-geom/9202026}}.

\bibitem{instanton}
A.~Klemm, {\it {Instanton}},  {\em
  \href{http://www.th.physik.uni-bonn.de/th/People/netah/cy/codes/inst.m}{http://www.th.physik.uni-bonn.de/th/People/netah/cy/codes/inst.m}}.

\end{thebibliography}\endgroup

\end{document}